\documentclass{nature}
\usepackage{amsmath,amssymb,amsfonts,latexsym,listings}
\usepackage{color}
\usepackage{graphicx}

\DeclareMathOperator*{\argmin}{arg\,min}

\newcommand{\captionfonts}{\footnotesize \baselineskip 12pt}
\makeatletter  
\long\def\@makecaption#1#2{%
  \vskip\abovecaptionskip
  \sbox\@tempboxa{{\captionfonts #1: #2}}%
  \ifdim \wd\@tempboxa >\hsize
    {\captionfonts #1: #2\par}
  \else
    \hbox to\hsize{\hfil\box\@tempboxa\hfil}%
  \fi
  \vskip\belowcaptionskip}
\makeatother   

\makeatletter
\renewcommand{\fnum@figure}{\textbf{Fig.~\thefigure}}

\renewcommand \thesection{~}
\renewcommand{\section}{\@startsection{section}{1}{0mm}
{0mm}%
{2mm}{\large\bfseries}}%

\renewcommand{\subsection}{\@startsection{subsection}{1}{\z@}
{\baselineskip}%
{-2mm}{\bf}}%
\makeatother

\title{{\normalsize\it  Nature Communications {\bf \emph{4}}, 1942 (2013)}\\ \\
Realistic Control of Network Dynamics}

\author{
Sean P. Cornelius$^1$,
William L. Kath$^{2,3}$ \&
Adilson E. Motter$^{1,3,\#}$
}

\begin{document}
\newcommand{\vect}[1]{\mathbf{#1}}
\newcommand{\x}{\vect{x}}
\newcommand{\xm}[1]{\x^{(#1)}}
\newcommand{\bxm}[1]{\mathbf{\xm{#1}}}
\newcommand{\dx}{\vect{\delta x}}
\newcommand{\Dx}{\vect{\Delta x}}
\newcommand{\mthit}[2]{#1^{(#2)}}
\newcommand{\target}{\x^*}
\newcommand{\eps}{\epsilon}
\newcommand{\M}{\mathbf{M}}

\baselineskip16pt
\setlength{\parskip}{0.05in}

\maketitle

\begin{affiliations}
 \item Department of Physics and Astronomy, Northwestern University, Evanston, IL 60208, USA.
 \item Engineering Sciences and Applied Mathematics, Northwestern University, Evanston, IL 60208, USA.
 \item Northwestern Institute on Complex Systems, Northwestern University, Evanston, IL 60208, USA.
\end{affiliations}

\vspace{-7mm}\noindent $^\#$ E-mail: \texttt{motter@northwestern.edu} (Corresponding author).
\vspace{5mm}

{\noindent \bf
The control of complex networks is of paramount importance in areas as diverse as ecosystem management, emergency response, and cell reprogramming. A fundamental property of networks is that perturbations to one node can affect other nodes, potentially causing the entire system to change behavior or fail. Here, we show that it is possible to exploit the same principle to control network behavior. Our approach accounts for the nonlinear dynamics inherent to real systems, and allows bringing the system to a desired target state even when this state is not directly accessible due to constraints that limit the allowed interventions. Applications show that this framework permits reprogramming a network to a desired task as well as rescuing networks from the brink of failure---which we illustrate through the mitigation of cascading failures in a power-grid network and the identification of potential drug targets in a signaling network of human cancer.
} \\

\noindent Complex systems such as  power grids, cellular networks, and food webs are often modeled as networks of dynamical units. In such complex networks, a certain incidence of  perturbations and the consequent impairment  of the function of individual units---whether power stations, genes, or species---are largely unavoidable in realistic situations.  While local perturbations may only rarely disrupt a complex system, they can propagate through the network as the system accommodates to a new equilibrium. This in turn often leads to system-wide re-configurations that can manifest themselves as genetic diseases\cite{motter_bioessays_2010,barabasi_review_2011}, power outages\cite{Carreras_ieee_2004,Buldyrev_nature_2010}, extinction cascades\cite{Pace_trends_1999,Scheffer_nature_2001}, traffic congestions\cite{helbing_rmp_2001,Vespignani_science_2009}, and other forms of large-scale failures\cite{May_nature_2008,Haldane_nature_2011}.

A fundamental characteristic of most large complex networks, both natural and man-made, is that they operate in a decentralized way. On the other hand, such networks have generally either evolved or been engineered to inhabit stable states in which they perform their functions efficiently. The existence of stable states indicates that arbitrary initial conditions converge to a relatively small number of persistent states, which are generally not unique and can change in the presence of large perturbations. Because complex networks are decentralized, upon perturbation the system can spontaneously go to a state that is less efficient than others available.  For example, a damaged power grid  undergoing a large blackout may still have other stable states in which no blackout would occur, but the perturbed system may not be able to reach those states spontaneously.  We suggest that many large-scale failures are determined by the convergence of the network to a ``bad'' state rather than by the unavailability of ``good'' states.

Here we explore the hypothesis that one can design physically admissible {\it compensatory perturbations} that can be used to direct a network to a desirable state even when it would spontaneously go to an undesirable (``bad'') state. An important precedent comes from the study of metabolic networks of  single-cell organisms, where perturbations caused by genetic or epigenetic defects can lead to nonviable strains.  The  knockdown or knockout of specific genes has been predicted to mitigate the consequences of such defects and often recover the ability of the strains to grow\cite{motter_msb_2008}. 
Another precedent comes from the study of food-web networks, where perturbations caused by human or natural forces can lead to the subsequent extinction of multiple species. Recent research predicts that a significant fraction of these extinctions can be prevented by the targeted suppression of specific species in the system\cite{sagar_natcom_2011}. These findings have  analogues in power grids, where perturbations caused by equipment malfunction/damage or operational errors can lead to large blackouts, but appropriate shedding of power can substantially reduce subsequent failures\cite{dobson_report_2010,anghel_ieee_2007}.   Therefore, the concept underlying our hypothesis is supported by recent research on physical\cite{Carreras_chaos_2002,motter_prl_2004}, biological\cite{motter_msb_2008,cornelius_pnas_2011}, and ecological networks\cite{sagar_natcom_2011}. The question we pose is whether compensatory perturbations can be systematically identified for a general network of dynamical units.

\section*{Results}
\subsection*{Control strategy for networks.\ }
Our solution to this problem is based on the insight that associated with each desirable state there is a region of initial conditions whose trajectories converge to it---the so-called ``basin of attraction'' of that state.  Given a network that is at (or will approach) an undesirable state, the conceptual problem is thus equivalent to identifying  a perturbation to the state of the system that can bring it to the attraction basin of the desired stable state (the {\it target} state). Once there, the system will evolve spontaneously to the target.  However, such perturbations must be physically admissible and are therefore subject to constraints---in the examples above, certain genes can be down-regulated but not over-expressed, the populations of certain species can only be reduced, and changes in power flow are limited by capacity and the ability to modify the physical state of the components. Under such constraints, the identification of a point within the target's basin of attraction is a highly nontrivial task.   

\begin{figure*}[th!]
\centering
\includegraphics[angle=0,width=12cm]{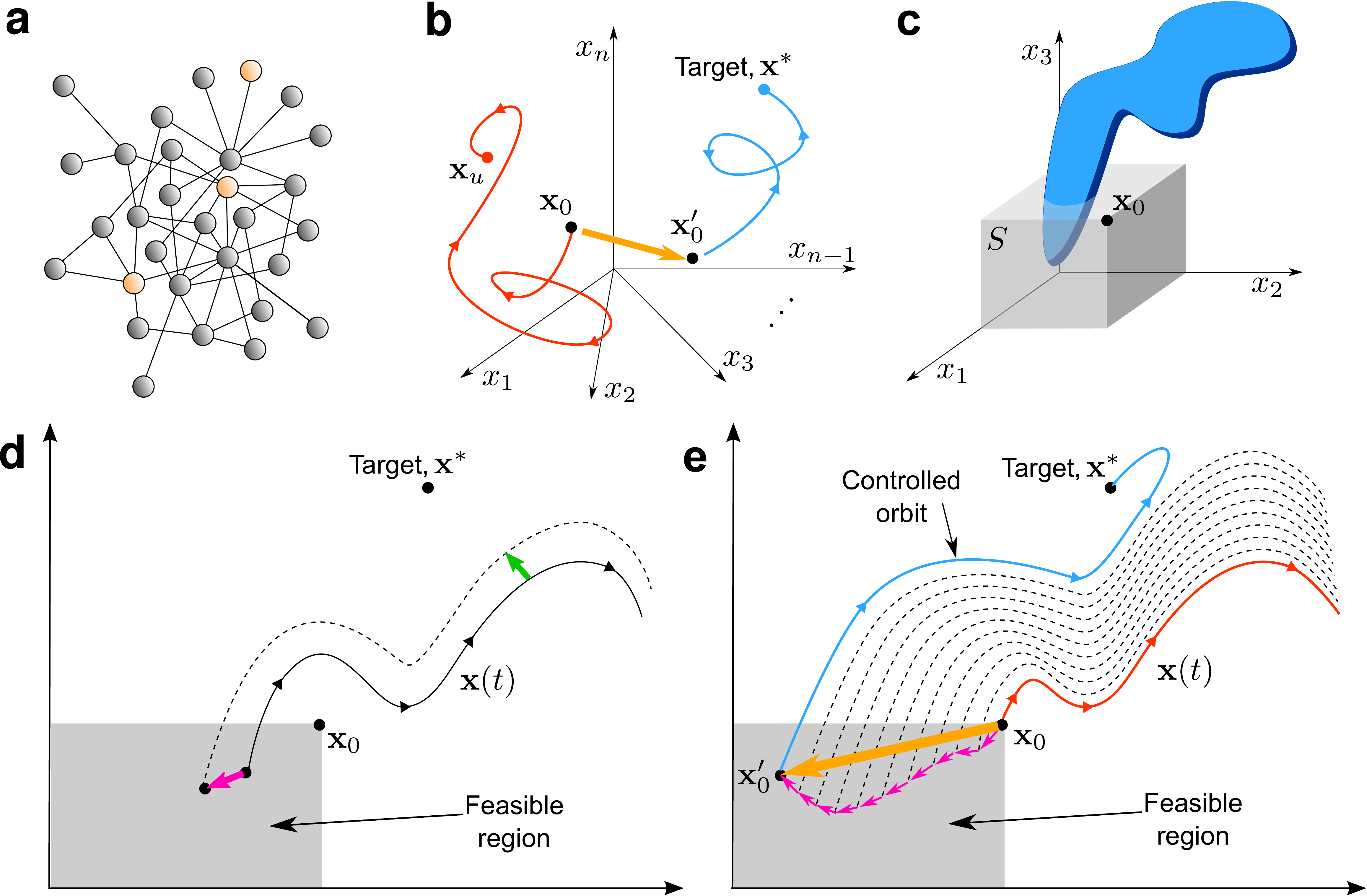}
\caption{{\bf Schematic illustration of the network control problem.}  ({\bf a}) Network portrait. The goal is to drive the network to a desired state by perturbing nodes in a {\it control set}---a set consisting of one or more nodes accessible to  compensatory perturbations. ({\bf b}) State space portrait. In the absence of control, the network at an initial state $\x_0$ evolves to an undesirable equilibrium $\x_\mathrm{u}$ in the $n$-dimensional state space (red curve). By perturbing the initial state (orange arrow), the network reaches a new state  $\x_0'$ that evolves to the desired target state $\target$ (blue curve). ({\bf c}) Constraints. In general, there will be constraints on the types of compensatory perturbations that one can make. In this example, one can only perturb three out of $n$ dimensions (equality constraints), which we assume to correspond to a thee-node control set, and the dynamical variable along each of these three dimensions can only be reduced  (inequality constraints). This results in a set of eligible perturbations, which in this case forms a cube within the three-dimensional  subspace of the control set.  The network is controllable if and only if the corresponding slice of the target's basin of attraction (blue volume) intersects this region of eligible perturbations (grey volume).  ({\bf d}, {\bf e})  Iterative construction of compensatory perturbations. ({\bf d}) A perturbation to a given initial condition (magenta arrow) results in a perturbation of its orbit (green arrow) at the point of closest approach to the target.  At every step, we seek to identify a perturbation to the initial condition that brings the closest-approach point of the orbit closer to the target. ({\bf e}) This process generates orbits that are increasingly closer to the target (dashed curves), and is repeated until a perturbed state  $\x_0'$  is identified that evolves to the target.  The resulting compensatory perturbation  $\x_0 \rightarrow \x_0'$ (orange arrow) brings the system to the attraction basin of the target without any {\it a priori} information about its location, and allows directing the network to a state that is not directly accessible by any eligible perturbation. \label{schematic_fig}}
\end{figure*}

Figure \ref{schematic_fig}a-c illustrates the problem that we intend to address. The dynamics of a network is best studied in the state space, where we can follow the time evolution of individual trajectories and characterize the stable states of the whole system.  Figure \ref{schematic_fig}a represents a network that would spontaneously go to an undesirable state, possibly due to an external perturbation, and that we would like to bring to a desired stable state by intentionally perturbing at most three of its nodes (highlighted).  Figure \ref{schematic_fig}b shows how this perturbation, changing the state of the system from $\x_0$ to $\x_0'$, would lead to an orbit that asymptotically goes to the target state.  As an additional constraint, suppose that the activity of the nodes is non-negative and can only be reduced (not increased) by this perturbation. Then, in the subspace corresponding to the nodes that can be perturbed, the set of points $S$ that can be reached by eligible perturbations forms a cubic region, as shown in  Fig.~\ref{schematic_fig}c.  The target state itself is outside this region (and, in fact, assumed to be outside the subspace of the three accessible nodes), meaning that it cannot be directly reached by {\it any} eligible perturbation. However, its basin of attraction may have points inside the region of eligible perturbations (Fig.~\ref{schematic_fig}c), in which case the target state can be reached by bringing the system to one of these points; once there, the system will spontaneously evolve toward the target state. This scenario leads to a very clear conclusion: a compensatory perturbation exists if and only if the region formed by eligible perturbations overlaps with the basin of attraction of the target.

However, there is no general method to identify basins of attraction (or this possible overlap) in the high-dimensional state spaces typical of complex networks (even though the desired stable states themselves are usually straightforward to identify). Despite significant advances, existing numerical techniques are computationally prohibitive and analytical methods, such as those based on  Lyapunov stability theorems, offer only rather conservative estimates and are not yet sufficiently developed to be used in this context\cite{zhong2009,Genesio1985,Kaslik2005}. Accordingly, our approach does not assume any information about the location of the attraction basins and addresses a problem that cannot be solved by existing methods from control theory, optimization, or network theory (Supplementary Discussion). 

\subsection*{Systematic identification of compensatory perturbations.\ }
The dynamics of a complex network can often be represented by a set of coupled ordinary differential equations. 
We thus consider an $N$-node network whose  $n$-dimensional dynamical state $\x$ is governed by 
\begin{equation} \label{system} 
\frac{d\x}{dt} = \vect{F}(\x).
\end{equation}
We focus on models of this form because of their widespread use and availability in modeling real complex networks. However, with minor modification, the approach we develop remains effective  in situations that, due to stochasticity and/or  parameter uncertainty, depart from idealized deterministic models (Supplementary Discussion, Supplementary Figs.~S1-S3).

The example scenario we envision is the one in which the network has been 
perturbed at a time prior to $t_0$, bringing it to a state $\x_0=\x(t_0)$ in the attraction basin $\Omega(\x_\mathrm{u})$ of an undesirable state $\x_\mathrm{u}$. 
We seek to identify a judiciously-chosen perturbation $\x_0 \rightarrow \x_0'$ to be implemented at time $t_0$ so that $ \x_0'$ belongs to the basin of attraction $\Omega(\target)$ of a desired state  $\target$.  For simplicity, 
we assume that $\x_\mathrm{u}$ and $\target$ are fixed points, although the approach we develop extends to other types of attractors. In the absence of any constraints it is always possible to perturb $\x_0$ such that  $\x_0' \equiv \target$. However, as discussed above, usually only  constrained compensatory perturbations are allowed in real networks. These constraints encode practical considerations and  often take the form of mandating no modification to certain nodes, while limiting the extent and direction of the changes in others. The latter is a consequence of the relative ease of removing versus adding resources to real systems. 
We thus assume that the constraints on the eligible perturbations can be represented  by vector expressions of the form
\begin{equation} \label{constraints}
\mathbf{g}(\x_0, \x_0') \leq {\bf 0} \,  \mbox{ and } \, \mathbf{h}(\x_0, \x_0') = {\bf 0},
\end{equation}
where the equality and inequality are interpreted to apply component-wise. We propose to construct compensatory perturbations iteratively from small perturbations, as shown in Fig.~\ref{schematic_fig}d-e. Given a dynamical system in the form (\ref{system}) and an initial state $\x_0$ at time $t_0$, a small perturbation $\delta \x_0$  evolves in time according to $\delta \x(t)={\mathbf M}(\x_0, t)\cdot \delta \x_0$. The matrix ${\mathbf M}(\x_0, t)$ is the solution of the variational equation $d{\mathbf M}/dt= D{\mathbf F}(\x)\cdot {\mathbf M}$ subject to the initial condition ${\mathbf M}(\x_0, t_0)={\bf 1}$. We can use this transformation to determine the perturbation $\delta \x_0$ to the initial condition $\x_0$ (at time $t_0$) that, among the admissible perturbations, will render $\x(t_\mathrm{c}) + \delta\x(t_\mathrm{c})$ closest to $\target$ (Fig.~\ref{schematic_fig}d), where $t_c$ is the time of closest approach to the target along the orbit. Large perturbations can then be built up by iterating the process: every time $\delta \x_0$ is calculated, the current initial state, $\x_0'$, is updated to $\x_0'+\delta \x_0$, and a new $\delta \x_0$ is calculated starting from the new initial state (Fig.~\ref{schematic_fig}e).  A visualization of this iterative procedure in two dimensions can be found in the Supplementary Information (Supplementary Discussion, Supplementary Fig.~4, Supplementary Movie). 
Before proceeding, we stress that the compensatory perturbation---the only intervention to be actually implemented in the network---is defined by the sum of all $\dx_0$.

After each iteration, we test whether the new state reaches the target (Methods). If so,  a compensatory perturbation has been found and is given by  $\x_0'-\x_0$. Now, it may be the case that no compensatory perturbation can be found, e.g., if the feasible region $S$ does not intersect the target basin $\Omega(\target)$.  To account for this, we automatically terminate our search if the  system is not controlled within a sufficiently large number of  iterations (Methods). We have, however, benchmarked our approach using randomly-generated networks in which compensatory perturbations are known to exist under the given constraints (Supplementary Discussion). Our method succeeds in identifying them in 100\% of cases, thus providing confidence that the approach introduced here can indeed be used to control a network when it is theoretically possible to do so. We note that our approach is effective even when it has to cross multiple attraction basins (Supplementary Fig.~4b) and when the basin boundaries are complex (Supplementary Discussion, Supplementary Figs.~S5-S6).

The above benchmark also confirms the efficiency of our algorithm (Supplementary Figs.~S7-S8), for which the theoretical running time is $O(n^{2.5})$, where $n$ is the number of dynamical variables in the network (Supplementary Discussion). Computationally, this is not onerous, especially since the control of a network requires the identification of only one compensatory perturbation. This should be contrasted with the $O(\exp(n))$ time that would be required to determine the basin of attraction  at fixed resolution by direct sampling of the state space. 

\subsection*{Application to the identification of therapeutic interventions.\ }
We apply our approach to the identification of potential therapeutic targets in a form of human blood cancer (large granular lymphocytic leukemia) caused by the abnormal survival of certain white blood cells (cytotoxic T cells). These T cells are part of the immune system and are produced to attack infected or dysfunctional cells. Under normal conditions, once the compromised cells have been removed, a significant portion of the T cells undergo programmed cell death (apoptosis). The disease, T-LGL leukemia, results precisely from the failure of T cells to undergo apoptosis, and the consequent negative impact they have on normal cells of the body\cite{Sokol2006}. The identification of potential therapeutic interventions is important since, at present, there is no curative treatment for this disease.

\begin{figure*}[th!]
\centering
\includegraphics[width=12.0cm]{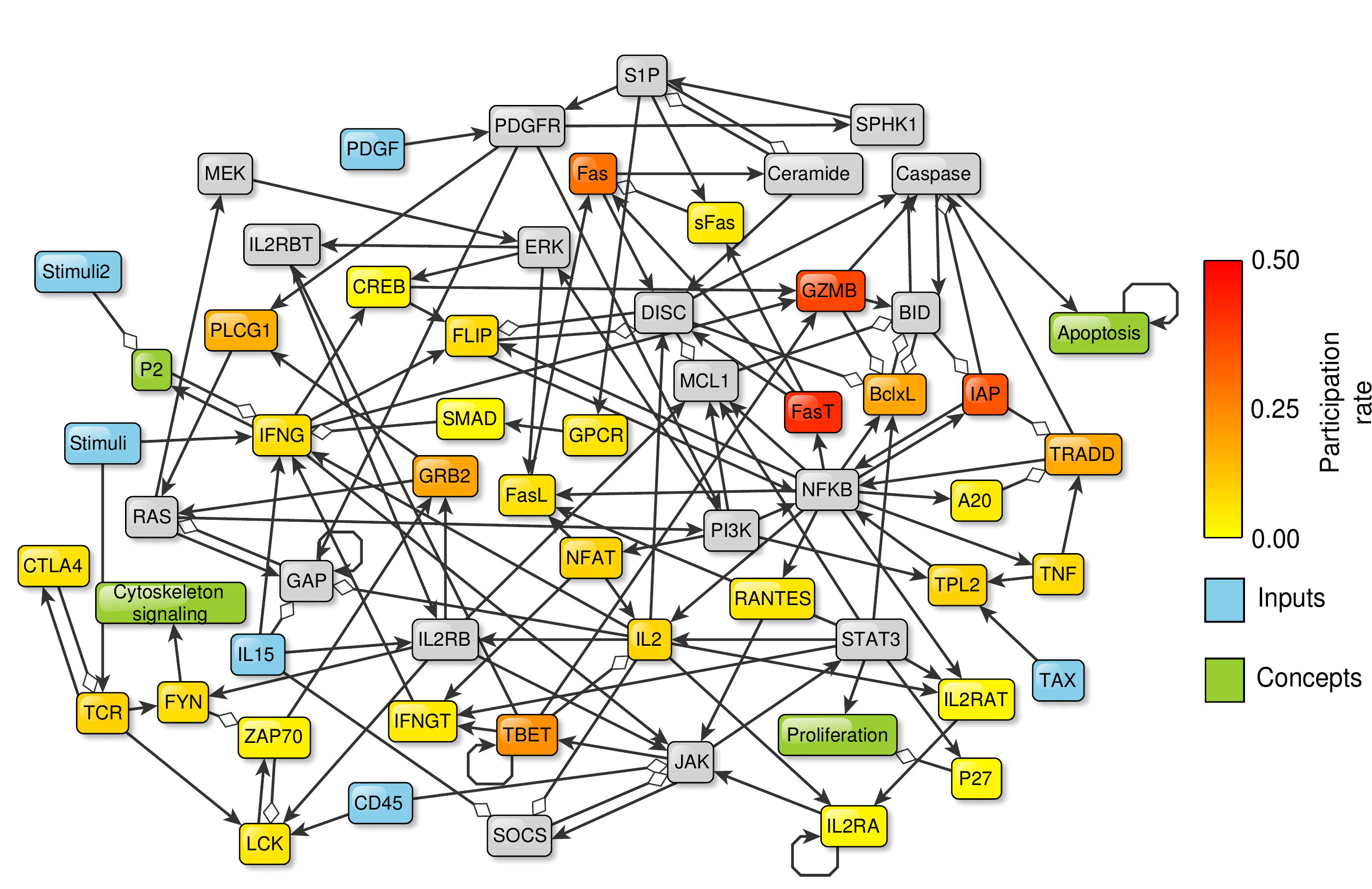}
\caption{{\bf T cell survival signaling network governing the development of T-LGL leukemia.}
Conceptual nodes, input nodes, and previously-identified potential therapeutic targets
are shown in green, blue, and grey, respectively. The edges
represent interactions, with the arrowheads and diamonds corresponding to activation and inhibition, respectively. 
The inhibitory edges that exist between Apoptosis and all non-input nodes are not shown for clarity. The 31 nodes colored yellow-red represent proteins and transcripts considered in our search for novel therapeutic targets, and are color-coded based on the frequency with which they appear (participation rate) in the smallest control sets
that we identify to successfully direct the network from a pre-cancerous state to the attraction basin of the normal cell state. Given a compensatory perturbation $\x_0 \rightarrow \x_0'$ (such that $\x_0$ is a pre-cancerous state and $\x_0'$ lies in the basin of attraction of the normal cell state), we find small control sets by first sorting the $31$ nodes under consideration in decreasing order based on the amount they were perturbed, and then searching for a new compensatory perturbation involving only the first $k$ nodes in this list. Through bisection on the number $k$, we are able to quickly converge to the ``minimal'' control set (with respect to this ordering) that can be used to rescue the given pre-cancerous state. This procedure is remarkably effective at producing small control sets---the average size is $3.4$ (with s.d.\  $3.7$).
\label{tcell_network_fig}
}
\end{figure*}

To formulate the problem, we use the 60-node survival signaling network model of T cells reconstructed and validated in ref.~\citen{loughran_pnas_2008}, where nodes correspond to proteins, transcripts, inputs (e.g., external stimuli), and cellular concepts (e.g., apoptosis).  The state of each node is  represented by a continuous variable between 0 and 1 (Methods). According to this model, normal and  cancer states correspond to two different types of stable steady states.  Potential curative interventions are those that can bring the system from  a cancerous or pre-cancerous state (those in the attraction basin of a cancer state) to the attraction basin of the normal state, which leads to apoptosis.  Previous experimental and computational studies have identified 19 nodes in this network as promising targets for curative interventions based on single, permanent reversals of the corresponding (binary) gene or protein activity in the cancer state\cite{albert_plos_2011}. The question we pose is whether novel interventions exist among the remaining nodes in the network (potentially involving multiple nodes), and furthermore, whether they can be effective with only the {\it temporary}, one-time perturbations considered here.  Figure \ref{tcell_network_fig} shows the $19$ previously characterized targets (grey), $10$ nodes representing static inputs (blue) or concepts (green), and the remaining $31$ nodes (yellow-red) that we use to search for novel interventions.

\begin{figure*}[th!]
\centering
\includegraphics[angle=0,width=12.0cm]{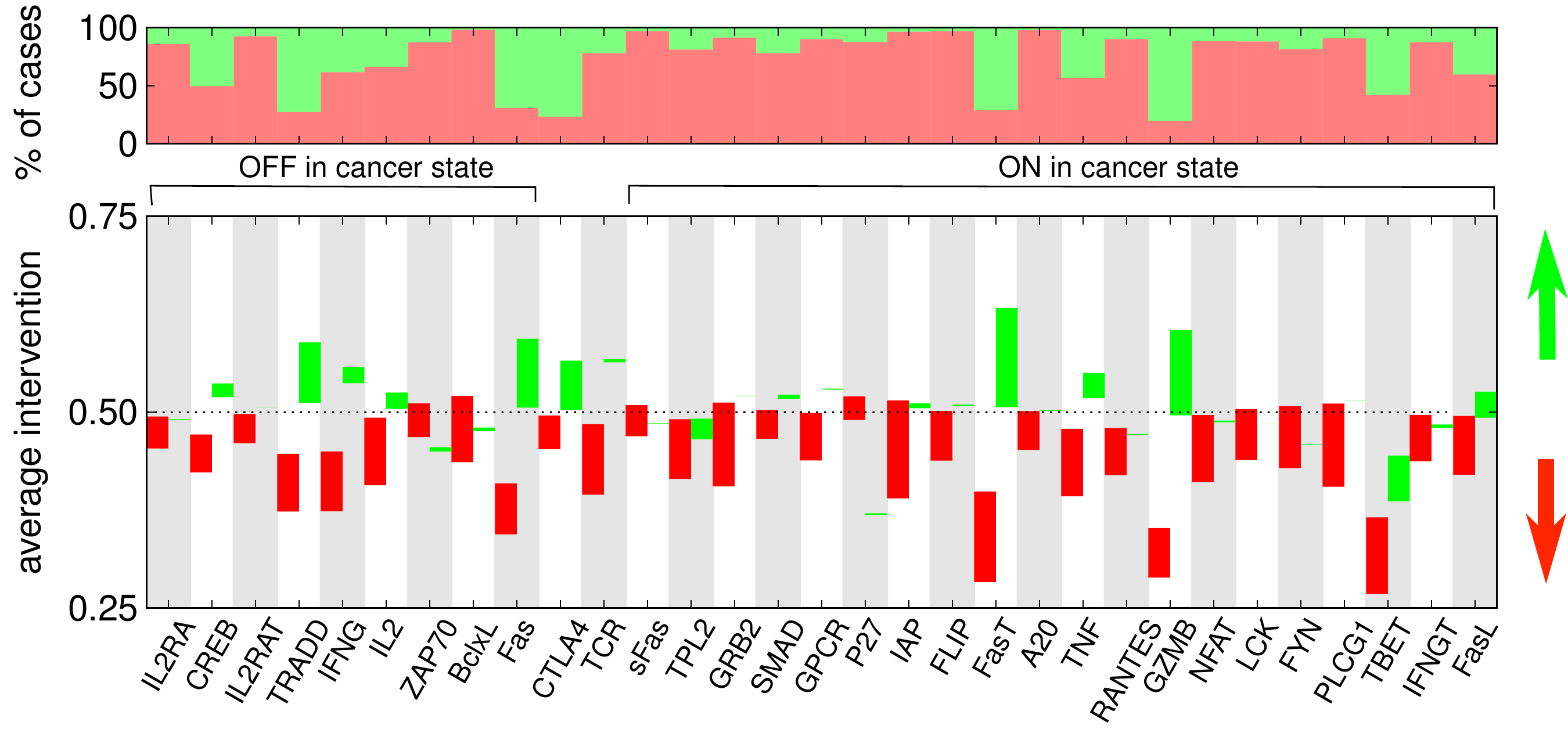}
\caption{{\bf Size and orientation of compensatory perturbations in the T cell survival signaling network.}
Each column corresponds to one of the 31 nodes under consideration as potential therapeutic targets, which
are ordered according to their predicted activity in the cancer state.
The data represent a sample of $10,000$ pre-cancerous network states, $6,731$ of which
are successfully rescued through compensatory perturbations identified by our approach.
The top panel shows the relative fraction of the successful interventions in which the activity of each individual node is increased (green)
versus decreased (red). The corresponding colors in the bottom panel represent the average pre-perturbation activity and the orientation and
size of the compensatory perturbation. Nodes are marked as either OFF or ON in the cancer state when their activity is $\approx 0$ or $\approx 1$ in that state, respectively (the only exceptions are the nodes CTLA4 and TCR, whose activity is $\approx 0.5$ in the cancer state). Remarkably, the interventions are 
such that a number of nodes are consistently perturbed toward, rather than away from, their activity levels in the undesirable (cancer) state.
\label{tcell_perturbations_fig}
}
\end{figure*}

We first allowed all of these $31$ accessible nodes to be perturbed, under the constraints that their state variables are kept within $0$ and $1$ and that the other nodes are not perturbed. Since it is important to consider intermediary cell states that  lead to the cancer state, we sought to identify compensatory perturbations for $10,000$ such states selected from a uniform sampling of the state space.  Of these,  $67 \%$ are successfully rescued using our approach. As shown in Fig.~\ref{tcell_perturbations_fig},  a number of striking patterns emerge in the interventions we found. Most nodes are consistently suppressed,
which may in part be attributed to the fact that all nodes other than Apoptosis are inactive in the target state, but this is also true for nodes that are inactive in the cancer state. In addition, there are several nodes whose activity is consistently enhanced,
despite the fact that they are active in the cancer state. These counterintuitive interventions are unlikely to be identified by simple inspection of the network or its stable states.

We can reduce the number of nodes that are perturbed by taking advantage of the reasonable expectation that the nodes that have been perturbed by the largest amount should dominate the membership of the smallest successful control sets (Fig.~\ref{tcell_network_fig}). Specifically, we find that we can rescue the same pre-cancerous states above with an average (standard deviation) of only $3.4$  ($3.7$) nodes. These interventions involve a small number of genes but at the same time are multi-target,  which is desirable given that the cure for currently incurable diseases is believed to reside in the coordinated modulation of multiple cellular components\cite{hopkins_08}. 
Such interventions are prohibitively difficult to identify experimentally by exhaustive search in the absence of computational predictions such as ours. Moreover, there is a high degree of overlap between these reduced control sets (Fig.~\ref{tcell_network_fig}), with nodes GZMB and FasT participating in nearly half of them. These nodes, and other frequently-occurring nodes such as IAP and Fas, are attractive candidates for experimental verification. Some of these genes work in tandem, with control sets formed by FasT and Fas alone predicted to rescue over $13\%$ of all cases. Our analysis suggests that interventions can be effective even if they are {\it temporary}, which, because they can be more easily implemented pharmacologically, are preferable to potential therapies based on permanent changes to a node state.

\subsection*{Reprogramming in associative memory networks.\ }
\begin{figure*}[th!]
\centering
\includegraphics[width=13.0cm]{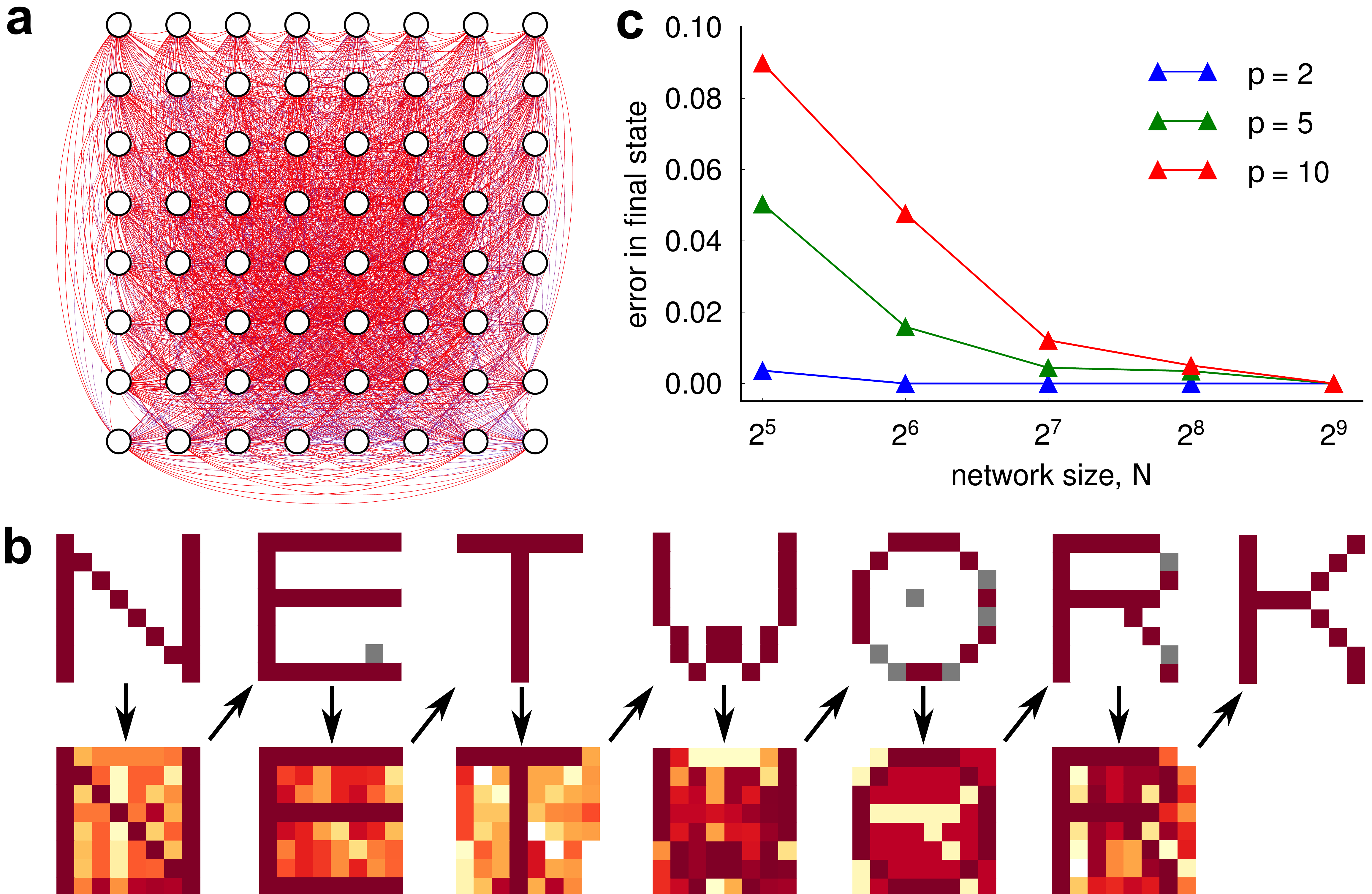}
\caption{{\bf Control in an associative-memory network.} ({\bf a}) Wiring diagram of a network of $N=64$ oscillators storing patterns representing the seven letters in the word ``NETWORK'', where red (blue) lines denote connections of positive (negative)  weight. ({\bf b}) Examples of transitions between memorized patterns induced by compensatory perturbations. Taking an initial state $\boldsymbol{\theta}_0$ corresponding to a letter in the word ``NETWORK'', we attempt to find a perturbation $\boldsymbol{\theta}_0 \rightarrow \boldsymbol{\theta}_0'$ (downward arrows) that then causes the network to spontaneously transition to the next letter under time evolution (diagonal arrows). Each oscillator is color-coded based on its angular distance from oscillator 1 (panel {\bf a}, upper left), while  the errors between the final state that is actually reached and the state that was targeted are indicated in grey. In each of the six cases, the control procedure successfully brings the system to the target or to a visually similar stable state with few such errors. ({\bf c}) Similar analysis shows that these errors become negligible as the  size of the network is increased relative to the number of stored patterns. The curves indicate the average  error (measured as the fraction of mismatched pixels)  between the target state and the final stable state actually reached using our control procedure. This average error is shown as a function of the network size $N$, for networks storing $2$ (blue), $5$ (green), and $10$ (red) random patterns, with each of the $N$ pixels having equal probability of being $\pm 1$. The coupling strength $\varepsilon$ is $0.2$, $0.4$, and $0.8$, respectively. Every point represents a set of $1,000$ independent network realizations, each sampled once, where the initial and target states are taken at random among the stored patterns.
\label{neural_fig}}
\end{figure*}

In an associative-memory network,  each memorized pattern is encoded as an attractor. An important problem in this context concerns the identification of constrained perturbations that cause the network to transition from a given pattern to a different specific pattern.  To illustrate this problem, we consider a model of associative memory consisting of $N$ identical coupled  
oscillators\cite{nishikawa_prl_2004},
\begin{equation} \label{neural_net}
\frac{d\theta_i}{dt} = \displaystyle \sum_{j=1}^N C_{ij} \sin(\theta_j - \theta_i) + \frac{\varepsilon}{N} \displaystyle \sum_{j=1}^N \sin 2(\theta_j - \theta_i),
\end{equation}  
where  $\theta_i$ is the phase variable of oscillator $i$,  $C_{ij}$ are the elements of the interaction matrix,  and $\varepsilon > 0$ 
is the strength of the second-order coupling term.  Up to translation of all oscillators by a constant phase, system  (\ref{neural_net}) has $2^N$ fixed points, corresponding to the phase-locked solutions in which $\lvert \theta_j - \theta_i \rvert = 0$ or $\lvert \theta_j - \theta_i \rvert = \pi$ for every 
$i$ and $j$. The attractors in this system consist of all such fixed points that are stable\cite{nishikawa_prl_2004}. This way, each asymptotic state of the network is identified 
uniquely with a binary pattern. In order to preferentially stabilize the desired states, the network is wired according to Hebb's learning rule $C_{ij} = \frac{1}{N} \sum_{\mu=1}^p \xi^\mu_i \xi^\mu_j$, where  $\boldsymbol{\xi}^\mu = (\xi^\mu_1, ..., \xi^\mu_N)$ with $\xi^\mu_i = \pm 1$ $(i=1,..,N, \mu=1,...,p)$ is the set of $p$ binary input patterns of length $N$ to be stored\cite{Hoppensteadt97}. As an example, we consider a network of size $N = 64$, for  $\varepsilon = 0.8$,  storing  $p=7$ patterns that represent the letters of the word ``NETWORK''. The resulting network is depicted in Fig.~\ref{neural_fig}a.  

We seek to identify perturbations that induce transitions between the memorized patterns while only changing oscillators representing ``off'' pixels, thereby requiring the existing pattern to be preserved. Figure \ref{neural_fig}b shows the results for initial/target state pairs corresponding to consecutive letters of ``NETWORK''.  In every case, the constraints on the eligible perturbations forbid reaching the target state directly. Nonetheless, in every case the control procedure succeeds in identifying a perturbed initial state (bottom row) that spontaneously evolves to the target or to a similar pattern with a small number of binary errors (grey)---and which is expected to become smaller in larger networks (Fig.~4c). Thus,  even if the basin of attraction of the target state cannot be reached by \emph{any} eligible perturbation, it may nonetheless be possible to drive the network to a similar state using our control procedure. 

\subsection*{Control of desynchronization instabilities in power-grid networks.\ }
\begin{figure*}[++th!]
\centering
\includegraphics[width=13.0cm]{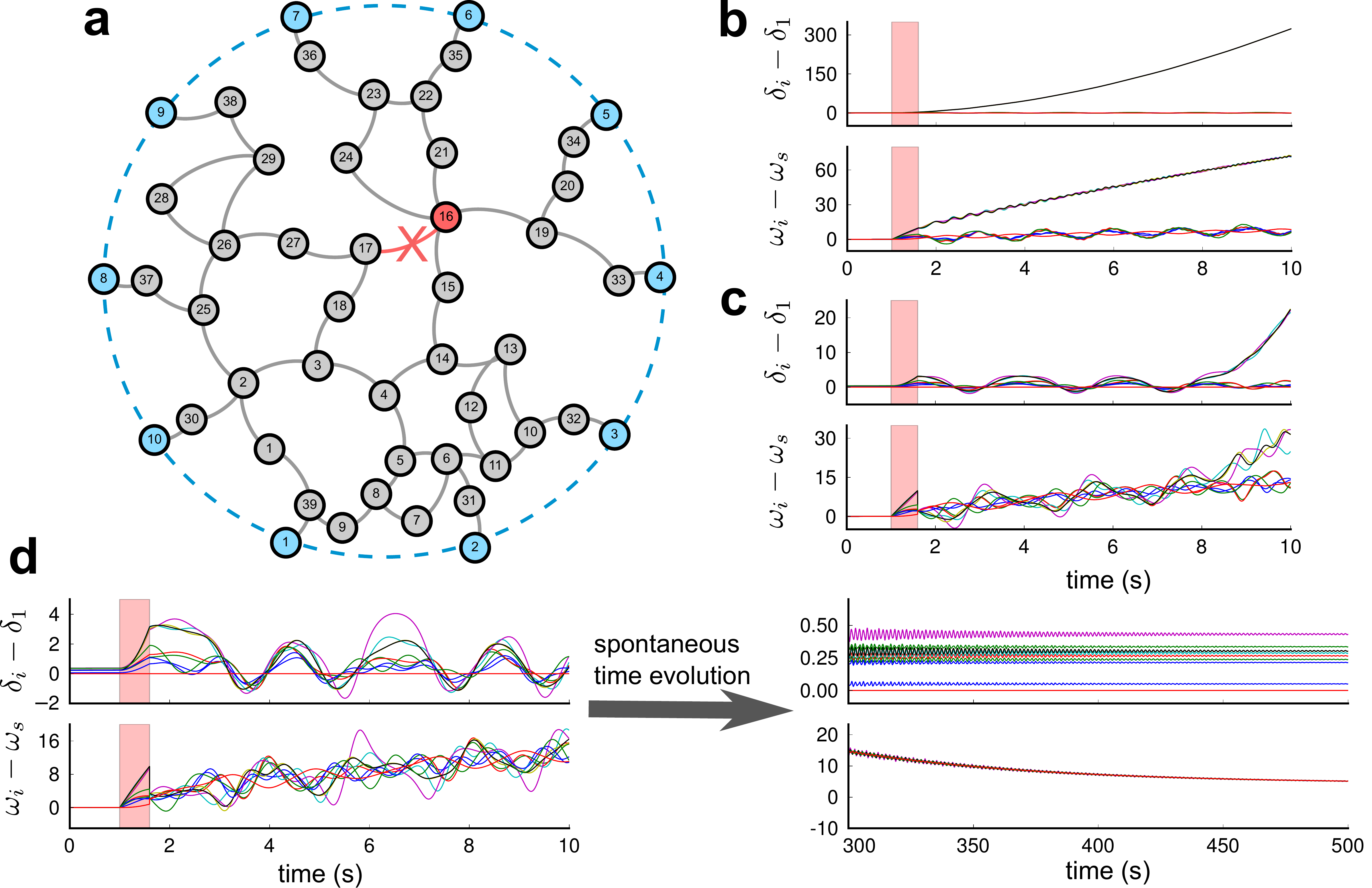}  
\caption{{\bf Control of the New England power-grid test system following a fault.} ({\bf a}) Schematic diagram of the network. The generators (the $N = 10$ dynamical nodes in the network) are highlighted in blue, and the non-generator nodes appear in grey.  The simulated fault is on the line connecting nodes 16 and 17 (red). It consists of short-circuiting  the end 16 of the line with the ground for the period $1.0-1.6$ s and subsequently removing the affected line from the network.  ({\bf b}-{\bf d}) Dynamics of the generators, characterized by the phases  $\delta_i$ (upper panels) and angular frequencies $\omega_i$ (lower panels): ({\bf b}) without any control perturbation, ({\bf c}) with a naive intervention based on resetting the generators' frequencies to the frequency of the target state, and ({\bf d}) with a compensatory perturbation identified by our control procedure. The fault induces a desynchronization ({\bf b}), which is not remediated by the naive intervention ({\bf c}), but the iterative control procedure  identifies a configuration of generator frequencies that maintains bounded swings in the short term ({\bf d}, left), and ultimately causes the system to evolve to the new synchronous (target) state ({\bf d}, right). This simplified example was chosen to have very large frequency deviations and transient period to facilitate visualization. In a realistic setting, the interventions can be implemented by tuning the damping of the generators. 
\label{new_england_fig}}
\end{figure*}

In the design and operation of power-grid networks, an important consideration is the ability of the power generators to 
maintain synchrony following perturbations\cite{anderson_book, takashi2013}. Desynchronization instabilities have in fact been implicated in cascading failures
underlying major recent blackouts\cite{Andersson2003}. The state of the system is assumed to be determined by the swing 
equation, 
\begin{equation}
\frac{2H_i}{\omega_s}\frac{d\omega_i}{d t}={P_{\text m}}_i - {P_{\text e}}_i,\;\;\; \frac{d\delta_i}{dt}= \omega_i, \,\,\,\,  i=1,\dots, N,
\label{b1-eq3}
\end{equation} 
where $N$ is the number of generators in the network and $\delta_i$ and $\omega_i$ are the phase and angular frequency of generator $i$, respectively. The constant $H_i$ is the inertia parameter of the generator, ${P_{\text m}}_i$ is the mechanical input power from the generator, and ${P_{\text e}}_i=-D_i\omega_i+\sum_{j=1}^{N}[D'_{ij}\sin(\delta_j-\delta_i)+D''_{ij}\cos(\delta_j-\delta_i)]$ is the power demanded of the generator by the network\cite{susuki_ieee_2009}. 
The network structure and impedance parameters are incorporated into the matrices $D'=(D'_{ij})$ and $D''=(D''_{ij})$, and the damping is accounted for by the coefficient $D_i$. In equilibrium, ${P_{\text m}}_i={P_{\text e}}_i$ and all generators operate in a synchronous state, characterized by  $\omega_1=\omega_2=\dots=\omega_N$. We illustrate our control procedure  on the New England power-grid model\cite{athay_ieee_1979}, which operates at the nominal synchronization frequency $\omega_s = 2\pi \times 60\,$rad s$^{-1}$ and consists of 10 generator nodes,  $39$ load nodes, and $46$ transmission lines (Fig.~\ref{new_england_fig}a). We implement this simple model for the parameter values given in ref.~\citen{susuki_ieee_2009}.

For an initially steady state solution determined by power flow calculations, we simulate single-line faults caused by short circuits  to the ground for a period of $0.6$s, during which the corresponding impedance is assumed to be very small  ($z = 10^{-9}$j)  and  at the end of  which the fault is cleared by disconnecting the line. Figure \ref{new_england_fig} shows one such  fault on the line connecting nodes $16$ and $17$ (Fig.~\ref{new_england_fig}a) and the corresponding time evolution of the $\delta_i$ and  $\omega_i$  for all generators in the network (Fig.~\ref{new_england_fig}b-d). By the time the fault is cleared, the generators have lost synchrony, and in the absence of any intervention, they continue accelerating away from one another (Fig.~\ref{new_england_fig}b). Nonetheless, the perturbed network admits a stable steady-state synchronous solution characterized by a new set of generator phases and a synchronization frequency only slightly different from $\omega_s$. In asking whether loss of synchrony can be averted by an appropriate compensatory perturbation following the fault, for illustrative purposes we assume  that direct modification of the generator phases $\delta_i$ is prohibited and only perturbations to the generator frequencies $\omega_i$ are allowed. 

A naive approach would be to reset the generator frequencies to the corresponding values at the target state after the fault, but, for not accounting for the full $2N$ dimensions of the state space, this approach fails and the system still loses synchrony, albeit at a later time (Fig.~\ref{new_england_fig}c). Using our iterative network control procedure, however,  one can identify a post-fault intervention that maintains bounded generator oscillations in the short term (Fig.~\ref{new_england_fig}d, left), and eventually causes the perturbed network to evolve to the desired target state (Fig.~\ref{new_england_fig}d, right). Out of the $92$ possible single-line fault perturbations of the type described above, $43$ cause the perturbed network to evolve to an undesirable  final state in which the generators have lost synchrony. Of these, $27$ cases can be controlled under the constraints described above.  
In each of the cases in which our method fails to find a compensatory perturbation, naive heuristic interventions---specifically, resetting the generators identically to either the nominal frequency or the synchronization frequency at the target state---also fail, suggesting that these perturbed networks may be impossible to control under the given constraints.  

\section*{Discussion}
\noindent The dynamics of large natural and man-made networks are usually highly nonlinear, making them complex not only with respect to their structure but also with respect to their dynamics. 
Nonlinearity has been the main obstacle to the control of such systems, and this is well reflected in the state of the art in the field\cite{liu2011, vicsek2012}. Progress has been made in the development of algorithms for decentralized communication and coordination\cite{bullo2009}, in the manipulation of Boolean networks\cite{Shmulevich2009}, in network queue control  problems\cite{Meyn2008}, 
and in other complementary areas.  
Methods have been developed for the control of networks hypothetically governed by linear dynamics\cite{lin_ieee_1974}. But although linear dynamics may approximate an orbit locally, control trajectories are inherently nonlocal\cite{rio}; moreoever, linear dynamics does not permit the existence of the different stable states observed in real networks and does not account for basins of attraction and other global properties of the state space. These global properties are crucial because they underlie network failures and, as shown here, provide a mechanism for network control. 
This can be achieved under rather general conditions  by systematically designing compensatory perturbations that take advantage of the full basin of attraction of the desired state, thus capitalizing on (rather than being obstructed by) the nonlinear nature of the dynamics.  

Applications show that our approach is effective even when compensatory perturbations are limited to a small subset of all nodes in the network, and when constraints forbid bringing the network directly to the target state.
From a network perspective, this frequently leads to counterintuitive situations  
in which the compensatory perturbations are in an opposite direction from that toward the target state---for example, suppressing nodes that are already less active than at the target. These results are surprising in light of the usual interpretation that nodes represent ``resources'' of the network, to which we then intentionally (albeit temporarily) inflict damage with a compensatory perturbation.  The same holds true for the converse, as demonstrated in the T cell survival signaling network. There the goal is to induce cell death, which, counterintuitively, is often achieved through perturbation toward the (active) cancer state. From the state space perspective,  the reason for the existence of such locally deleterious (beneficial) perturbations that have globally beneficial (deleterious) effects is that the basin of attraction, being nonlocal, can extend to the region of feasible perturbations even when the target itself does not.  

We have motivated our problem assuming that the network is away from its desired equilibrium due to an external perturbation. In particular, as shown in our example of desynchronization failures in a power grid, our approach can be used for the real-time rescue of a network, bringing it to a desirable state before it reaches a state that can be temporarily or permanently irreversible.  We suggest that this can be important for the conservation of ecological systems and for the creation of self-healing infrastructure systems. On the other hand, as illustrated in our associative-memory example, our approach also applies to move the network from one stable state to another, thus providing a mechanism for ``network reprogramming''. 

As a broader context to interpret the significance of this application, consider the reprogramming of differentiated (somatic) cells from a given tissue  into a pluripotent stem cell state, which can then differentiate into cells of a different type of tissue.  The seminal experiments demonstrating this possibility involved continuous overexpression of specific genes\cite{Takahashi_cell_2006}, which is conceivable even under the hypothesis that cell differentiation is governed by the loss of stability of the stem cell state\cite{Huang_DevBiol_2007}. However, the recent demonstration that the same can be achieved by the temporary expression of few proteins\cite{kim_stem_2009} or transient administration of mRNA\cite{collins_2010} indicates that  the stem cell state may have remained stable (or metastable) after differentiation,  
allowing interpretation of the reprogramming process in the context of the interventions considered here.
While induced pluripotency is an example par excellence of network reprogramming, the same concept extends far beyond this particular system.  Taken together, our results provide a new foundation for the control and rescue of network dynamics and, as such, are expected to have implications for the development of smart traffic and power-grid networks, of ecosystems and Internet management strategies, and of new interventions to control the fate of living cells.

\section*{Methods}
\subsection*{Identification of compensatory perturbations.}\ 
We identify compensatory perturbations iteratively as follows. Given the current initial state of the network, $\x_0'$, we integrate the system dynamics
over a time window $t_0 \leq t \leq t_0 + T$  to identify the time of the orbit's closest approach to the target, $t_\mathrm{c} \equiv \argmin \lvert \target - \x(t) \rvert$.
We then integrate the variational equation up to this time to obtain the corresponding variational matrix, $\M(t_\mathrm{c})$, which maps
a small change $\dx_0$ to the initial state of the network to a change $\dx(t_\mathrm{c})$ in the resulting perturbed orbit at $t_\mathrm{c}$
according to $\dx(t_\mathrm{c}) = \M(t_\mathrm{c}) \cdot \dx_0$. This mapping is used to select an incremental perturbation $\dx_0$ to the current initial state that minimizes
the distance between the perturbed orbit and the target at time $t_\mathrm{c}$, subject to the constraints (\ref{constraints}) on the eligible perturbations
as well as additional constraints on $\dx_0$ to ensure the validity of the variational approximation (\emph{Constraints on incremental perturbations}, below). 
This selection is performed via a nonlinear optimization (\emph{Nonlinear optimization}, below). The initial condition is then updated according to 
$\x_0' \rightarrow \x_0'+\delta \x_0$, and we test whether the new initial state lies in the target's basin of attraction by integrating the system dynamics over a long time
$\tau$. If  the system's orbit reaches a small ball of radius $\kappa$ around $\target$ within this time, we declare success  and recognize
$\x_0 - \x_0'$ as a compensatory perturbation (for the updated $\x_0'$). If not, we calculate the time of closest approach of the new orbit and repeat the procedure, up
to a maximum number $I$ of iterations.

\subsection*{Constraints on incremental perturbations.}\ 
The incremental perturbation at the point of closest approach, $\delta \x(t_\mathrm{c})$, selected under constraints (\ref{constraints}) alone will generally have a nonzero component along a stable subspace of the orbit $\x(t)$, which will result in  $\delta \x_0$ larger than $\delta \x(t_\mathrm{c})$ by a factor of up to $O(\exp|\lambda_s^{t_\mathrm{c}}(t_\mathrm{c}-t_0)|)$, where $\lambda_s^{t_\mathrm{c}}$ is the finite-time Lyapunov exponent of the eigendirection corresponding to the eigenvalue of ${\mathbf M}(\x_0, t_\mathrm{c})$ with smallest-magnitude real part. In a naive implementation of this algorithm, to keep $\delta \x_0$ small for the linear transformation to be valid, the size of $\delta \x(t_\mathrm{c})$ would be negligible, leading to negligible progress. This problem is avoided by optimizing the choice of $\delta \x(t_\mathrm{c})$ under the constraint that the size of $\delta \x_0$ is bounded above. Another potential problem is when the perturbation causes the orbit to cross an intermediate  basin boundary before reaching the final basin of attraction. All such events can be detected by monitoring the difference between the linear approximation and the full numerical integration of the orbit, without requiring any prior information about the basin boundaries. Boundary crossing is actually not a problem because the closest approach point is reset at each iteration and, in particular, on the new side of the basin boundary. To assure that the method will make satisfactory progress at each iteration, we solve the optimization problem under the constraint that the size of $\delta \x_0$ is also bounded below, which means that we accept increments $\delta \x_0$ that may temporarily increase the distance from the target. 
These upper and lower bounds can be expressed as
\begin{equation}
\epsilon_0\le |\delta \x_0| \le \epsilon_1.
\label{constraints2}
\end{equation}
This can lead to $|\delta \x(t_\mathrm{c})|\gg\epsilon_1$ due to components along the unstable subspace, but in such cases the vectors
can be rescaled after the optimization.  
At each iteration, the problem of identifying a perturbation $\delta \x_0$ that incrementally moves
the orbit toward the target under constraints (\ref{constraints}) and (\ref{constraints2}) is then solved 
as a constrained optimization problem. To avoid back-and-forth oscillations, we require the inner product between two consecutive increments $\delta \x_0$ to be positive. 
The resulting iterative procedure behaves well as long as  $\lvert \tilde{\delta} \x (t_\mathrm{c}) - {\mathbf M} (\x_0,t_\mathrm{c}) \cdot \delta \x_0  \rvert= O(\lvert \delta \x_0\rvert^2)$, which can be
assured  by properly choosing $\epsilon_0$ and $\epsilon_1$,  where  $\tilde{\delta} \x (t_\mathrm{c})$ is the actual change in the orbit at $t_\mathrm{c}$ measured when the orbit is integrated anew at the subsequent iteration. In practice, the approach does not depend critically on very accurate forecasting of $\delta \x(t_\mathrm{c})$ at any single iteration so long as it moves the orbit closer to the target, and it is observed to be effective for a rather wide range of parameters $\epsilon_0$ and $\epsilon_1$.

\subsection*{Nonlinear optimization.}\
The optimization step of the iterative control procedure
consists of finding the small perturbation $\delta \x_0$
that minimizes the remaining distance between the target, $\target$,
and the system orbit $\x(t)$ at its time of closest approach, $t_\mathrm{c}$.
Constraints are used to define the admissible perturbations (\ref{constraints}) and also, 
as described in the previous paragraph, to limit the magnitude of $\delta \x_0$ (\ref{constraints2}).
The optimization problem to identify $ \delta \x_0$ can then
be  succinctly written as:
\begin{align} \label{nlp}
\textrm{min} & & \lvert \target - (\x(t_\mathrm{c}) + {\mathbf M}(\x_0', t_\mathrm{c}) \cdot \delta \x_0) \rvert &\\
\textrm{s.t.} & & \mathbf{g}(\x_0, \x_0' + \delta \x_0) & \leq 0 \label{ieq_constraint} \\
& & \mathbf{h}(\x_0, \x_0' + \delta \x_0) & =  0 \label{eq_constraint} \\
& & \eps_0 \; \leq \lvert \delta \x_0 \rvert \; & \leq \eps_1 \label{size_constraint} \\
& & \delta \x_0 \cdot \delta \x_0^{p} & \geq 0, \label{direction_constraint}
\end{align}
where (\ref{direction_constraint}) is enforced starting from the second iteration,  and
$\delta \x_0^{p}$ denotes the incremental perturbation from the previous iteration.
Formally, this is a nonlinear programming (NLP) problem, the solution of
which is complicated by the nonconvexity of the constraint (\ref{size_constraint}) (and possibly
(\ref{ieq_constraint}) and (\ref{eq_constraint})). Nonetheless, a number of algorithms have been developed for 
the efficient solution of NLP problems, among them sequential quadratic programming (SQP)\cite{sqp}.
This algorithm solves (\ref{nlp}) subject to (\ref{ieq_constraint})-(\ref{direction_constraint})
as the limit of a sequence of quadratic programming
subproblems, in which the constraints are linearized in each sub-step. For all calculations, we use
the SQP algorithm\cite{kraft} implemented in the SciPy scientific programming package (http://www.scipy.org/).
In all systems, we use dimensionless distances. In the case of the power-grid network, this is implemented by normalizing 
frequency by the target frequency, which further avoids disparate scales between the frequency and phase variables.
More generally, while the norm in (\ref{nlp}) may denote the usual Euclidean norm for most systems, there is nothing in our formulation
of the control procedure that prohibits optimizing closeness according to a different metric in a particular network, especially if 
the dynamical variables represent different quantities or are otherwise not of the same order.

\subsection*{Values of parameters.}\
For the values of the parameters $\kappa$, $\tau$, $I$, $\epsilon_0$, $\epsilon_1$, and $T$ used in the example applications of this paper, as well as criteria for choosing
their values in the general case, we refer the reader to the Supplementary Methods and Supplementary Table S1.\

\subsection*{T cell survival signaling network.}\
The network consists of 60 nodes, 54 of which are equipped with dynamics and represent the state of the network, while 6 are static input nodes (Stimuli, TAX, CD45, PDGF, Stimuli2, and IL15)\cite{loughran_pnas_2008, albert_plos_2011}. Following refs.~\citen{loughran_pnas_2008} and \citen{albert_plos_2011}, we set Stimuli, IL15, and PDGF at ON (one) and set 
TAX, CD45, and Stimuli2 at  OFF (zero) for all simulations.  We translate the Boolean network dynamics given in ref.~\citen{albert_plos_2011} into an equivalent continuous form using the method described in ref.~\citen{theis_bmc_2009}. The state variable $x_i$ representing the activity of each node is thus allowed to assume values  in the range $[0,1]$. The associated dynamics follows
\begin{equation}
\dot{x_i} = B_i(f(x_1),\ldots ,f(x_N)) - x_i.
\end{equation}
Here, $B_i$ is a continuous analog of the discrete Boolean update rule for node $i$, which would take the current
state (ON or OFF) of all nodes as an input and output the state of node $i$ at the next time step. The function $B_i$ is obtained via multilinear
interpolation of the associated logical function between the ``corners'' of the $N$-dimensional unit cube (in which the value
of each node is either $0$ or $1$). To capture the switch-like behavior observed in signaling circuits, the state of each node is passed through
a sigmoidal (Hill-type) function $f(x) = x^4/(x^4 + k^4)$ before it is used as an input to the continuous logical gates $B_i$. Nodes 
are considered to be ON (OFF) if the associated $x_i$ is significantly above (below) the threshold $k$, which we take to be $0.5$. The generation of the continuous model dynamics was done automatically with the software package Odefy\cite{theis_bmc_2010}. We observe three stable fixed points in the network. One fixed point corresponds to the normal cell state (the target state in our simulations), in which the node representing apoptosis is ON and all other dynamical nodes are OFF.  The two other fixed points  are biologically equivalent (differing only by node P2, which can be either ON or OFF) and correspond to the cancer state. The three attractors, as defined by the associated ON/OFF states of the individual nodes, are identical to those found in ref.~\citen{albert_plos_2011}.

\renewcommand{\refname}{References}

\subsection*{Acknowledgments}{~}\\
The authors thank Jie Sun for illuminating discussions.  This work was supported by NSF under Grant DMS-1057128, NCI under Grant 1U54CA143869-01, and a Northwestern-Argonne Early Career Investigator Award to A.E.M.

\subsection*{Author contributions}{~}\\
S.~P.~C., W.~L.~K., and A.~E.~M.\ conceived the study; S.~P.~C.\ performed the numerical experiments; S.~P.~C.\ and A.~E.~M.\ analyzed the data; and S.~P.~C.\ and A.~E.~M.\ wrote the manuscript. All authors approved the final version of the manuscript.

\subsection*{Additional information}{~}\\
The authors declare no competing financial interests. Correspondence should be addressed to A.E.M. (motter@northwestern.edu).

\pagebreak
\section*{Figures}

\newpage
\clearpage
\makeatletter
\renewcommand \thesection{S\arabic{section}.}
\renewcommand{\section}{\@startsection{section}{1}{0mm}
{\baselineskip}%
{\baselineskip}{\Large\bfseries}}%
\makeatother

\renewcommand{\thefigure}{S\arabic{figure}}
\renewcommand{\figurename}{{\bf Supplementary Figure }}
\makeatletter
\makeatletter \renewcommand{\fnum@figure}
{\figurename \thefigure}
\makeatother

\setcounter{figure}{0}
\setcounter{equation}{0}
\setcounter{table}{0}

\renewcommand{\thetable}{S\arabic{table}}
\renewcommand{\theequation}{S\arabic{equation}}
\renewcommand{\tablename}{{\bf Supplementary Table}}
\renewcommand{\listfigurename}{Supplementary Figures}

\begin{center}
{\bf \large Realistic Control of Network Dynamics} \\
\smallskip
{\it \large S.P. Cornelius, W.L. Kath, and A.E. Motter} \\
\smallskip
{\large Supplementary Information}
\end{center}

\tableofcontents

\newpage
\section*{Supplementary Figures}
\addcontentsline{toc}{section}{Supplementary Figures}

\begin{figure*}[th!]
\centering
\includegraphics[angle=0,width=16.0cm]{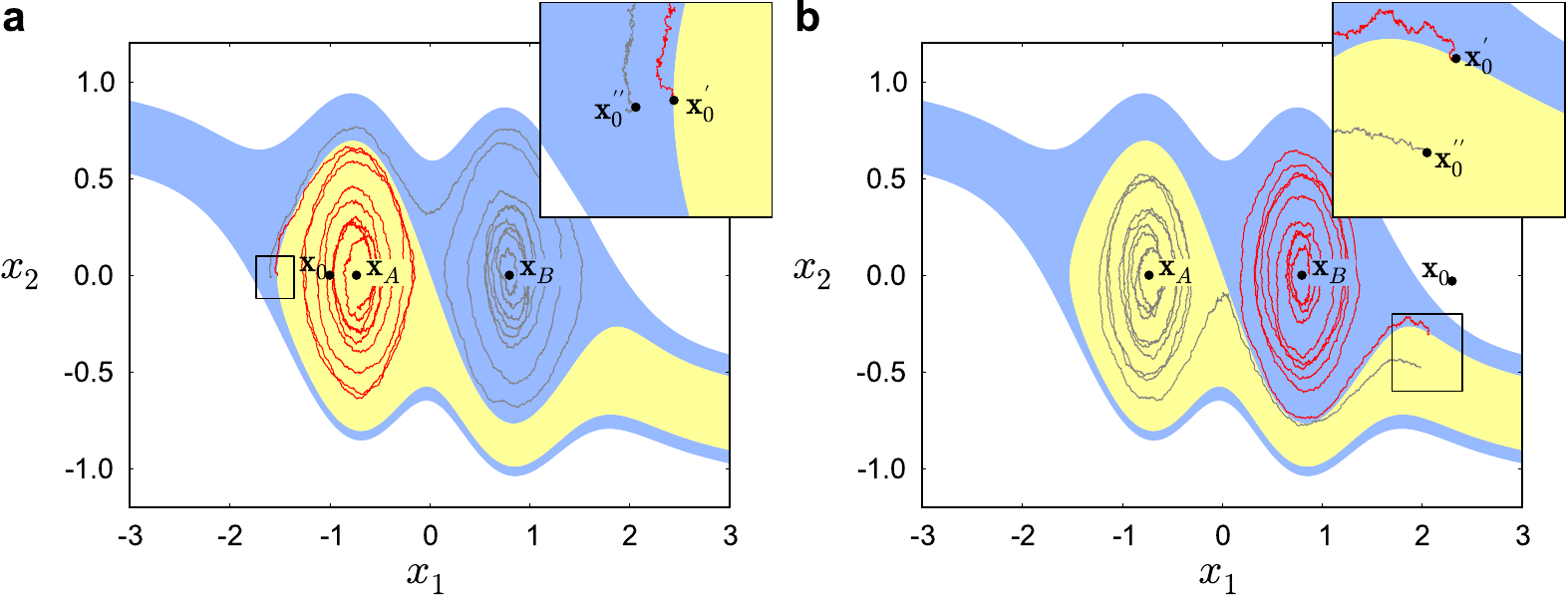}
\caption[]{{\bf Effects of stochasticity.} Scenarios ({\bf a}) and ({\bf b}) correspond to the control problems depicted in Supplementary Figures \ref{2d_example_fig}a and \ref{2d_example_fig}b, respectively,
where $\x_\mathrm{A}$ and $\x_\mathrm{B}$ are the attractors of the noiseless dynamics and yellow and blue their corresponding basins of attraction
in the deterministic case. The intervention $\x_0 \rightarrow \x_0'$ 
represents the compensatory perturbation which, in the absence of noise, results in an orbit that carries
the system to the target ($\x_\mathrm{B}$ and $\x_\mathrm{A}$, respectively). This example illustrates that when noise is added to the dynamics, however, the system
may fail to reach the target attractor, possibly approaching the other attractor instead (red curves). This can be attributed to the proximity of $\x_0'$ to the corresponding basin boundary. The situation can be remedied by making an additional perturbation $\x_0' \rightarrow \x_0''$ 
that places the system further inside the target basin of attraction, increasing the likelihood that the noisy orbit 
(grey) will reach the target. 
The noise strength is $s=0.03$ for all orbits pictured here. 
\label{noise_schematic_fig}}
\addcontentsline{toc}{subsection}{Supplementary Figure \thefigure}
\end{figure*}

\newpage
\clearpage
\begin{figure*}[th!]
\centering
\includegraphics[angle=0,width=14.0cm]{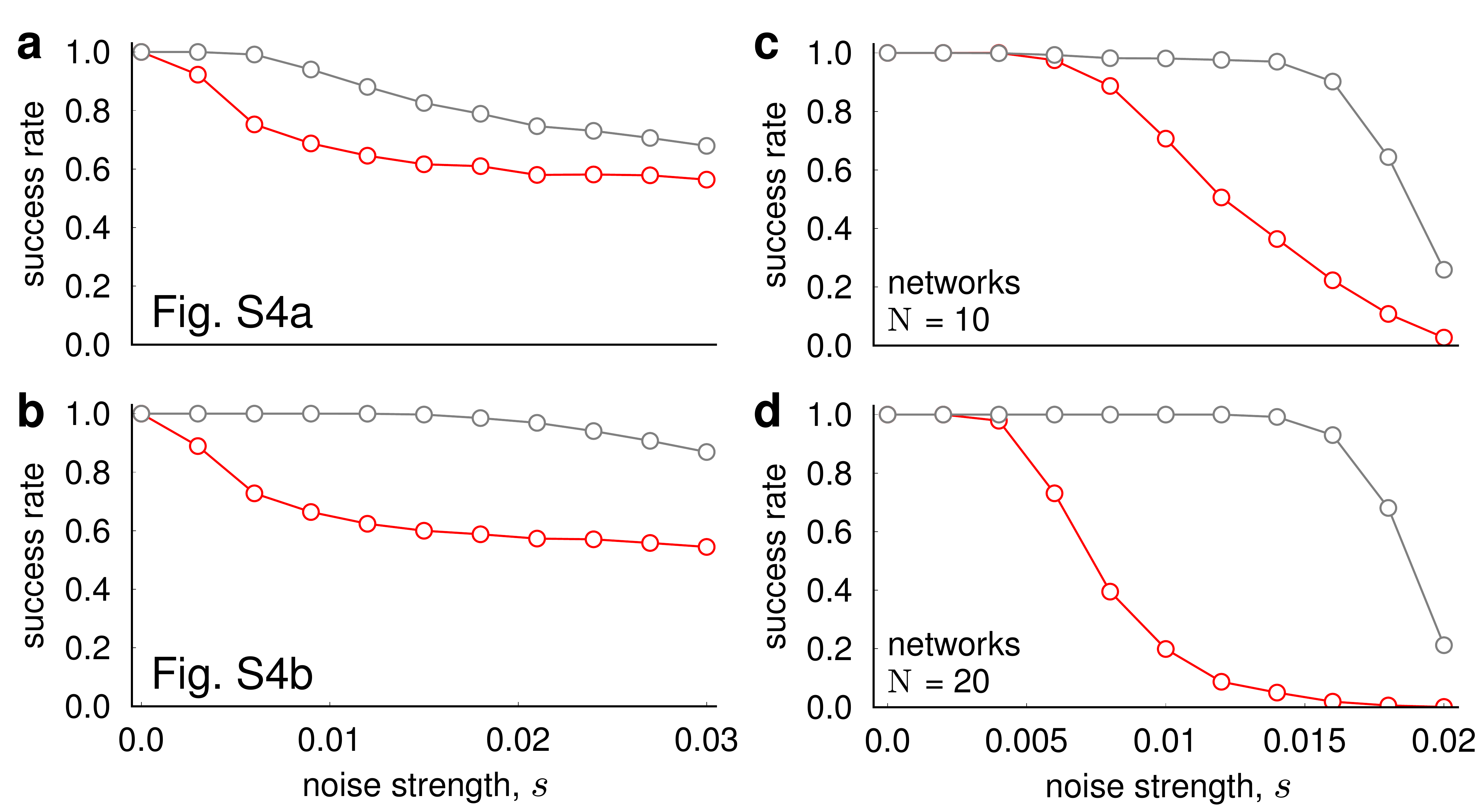}
\caption[]{{\bf Robustness of compensatory perturbations against noise.} The curves
indicate the success rate of compensatory perturbations predicted using deterministic models when noise 
of r.m.s.\ amplitude $s$ is added to the dynamics, for the compensatory perturbations as predicted using our original computational approach (red), and modified
compensatory perturbations that are systematically identified to be further inside the target basin of attraction (grey).
({\bf a}, {\bf b}) Mechanical example system, with the initial target states considered in Supplementary Figures \ref{2d_example_fig}a and \ref{2d_example_fig}b, respectively. ({\bf c}, {\bf d}) 
Ensemble of $100$ random networks and initial and target states as considered in Supplementary Figure \ref{random_networks_fig} for sizes $N=10$ and 20, respectively. In all scenarios, a point corresponds to 1,000 independent sample paths of the noisy dynamics. Each point in ({\bf a},  {\bf b}) represents 1,000 noisy orbits starting 
from the corresponding perturbed state, while each point in ({\bf c}, {\bf d}) represents 10 such orbits for each of the 100 network realizations of 
the given network size. In every case, the approach to the target can thus be insulated against noise with a slight modification of the 
compensatory perturbation procedure, thereby preserving its effectiveness. 
\label{noise_success_fig}}
\addcontentsline{toc}{subsection}{Supplementary Figure \thefigure}
\end{figure*}

\newpage
\clearpage
\begin{figure*}[th!]
\centering
\includegraphics[angle=0,width=12.0cm]{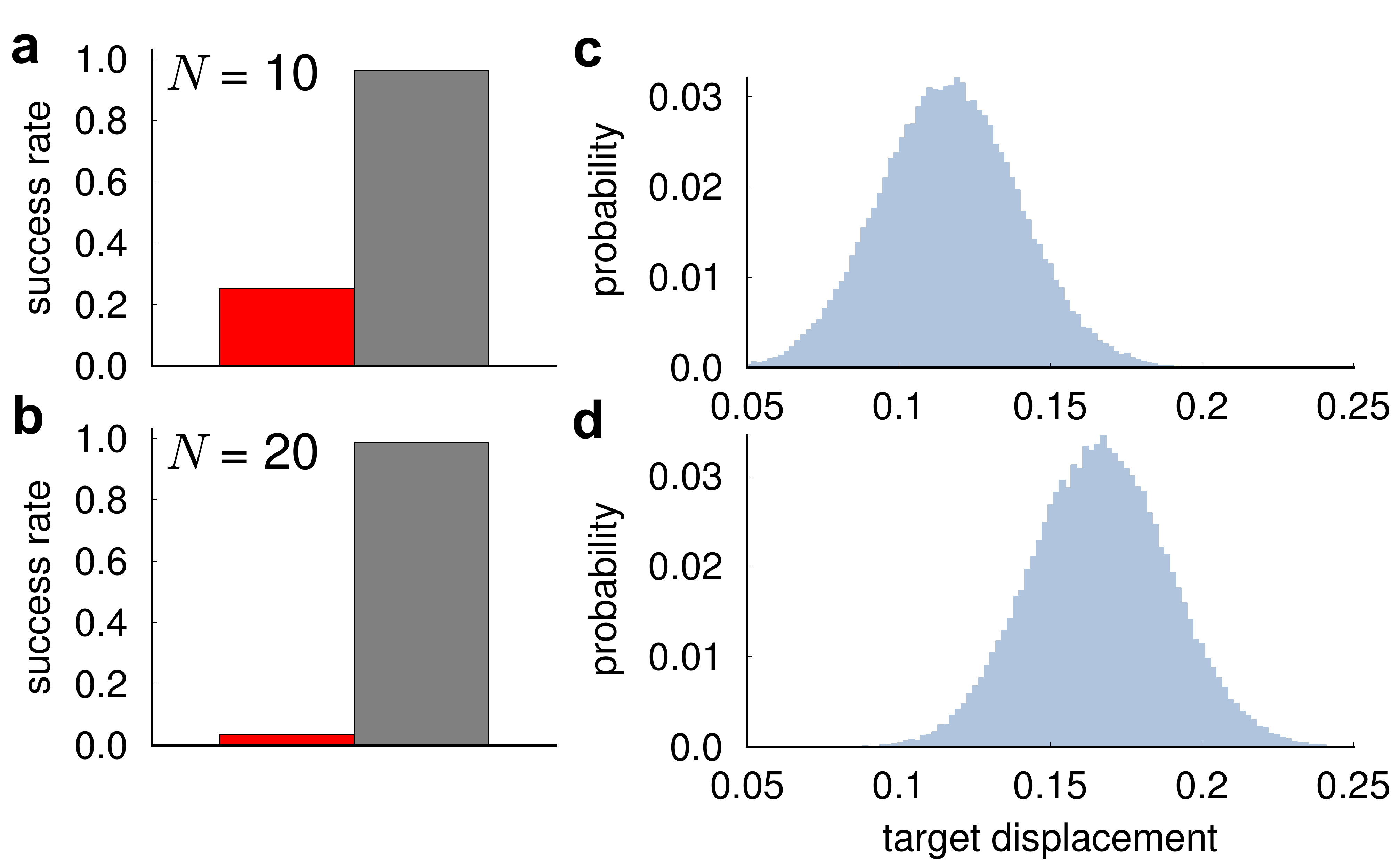}
\caption[]{{\bf Robustness of compensatory perturbations against parameter uncertainty.} ({\bf a}, {\bf b}) Bars 
denote the success rate of candidate compensatory perturbations in directing network to the target, for the compensatory perturbations 
predicted by our original computational procedure (red) and the modified compensatory perturbations as 
described in the Supplementary Discusison (grey). Each set of bars represents a sample of $1,000$ random networks and initial and target
states as considered in Supplementary Figure \ref{random_networks_fig}, and the success rate for each
of these networks is an average over $100$ randomized parameter sets in which we allow every parameter of the system to vary independently and uniformly 
within a range of $\pm 5\%$ about its nominal value. 
In both cases, the significantly higher success rate for the modified perturbations demonstrates that failure to
control the system is buffered against parameter uncertainty by driving the system further inside the target basin. 
({\bf c}, {\bf d}) Distribution of distances between the target state of the original network and the corresponding 
stable state of the networks with modified parameters considered in panels ({\bf a}) and ({\bf b}), respectively. The distances are 
normalized by the (common) equilibrium value of the dynamical units  in the target state $\vec{\x}_\mathrm{B}$ (compare with  Supplementary Fig.~\ref{random_networks_fig}a).
\label{network_params_fig}}
\addcontentsline{toc}{subsection}{Supplementary Figure \thefigure}
\end{figure*}

\newpage
\clearpage
\begin{figure*}[th!]
\centering
\includegraphics[width=16cm]{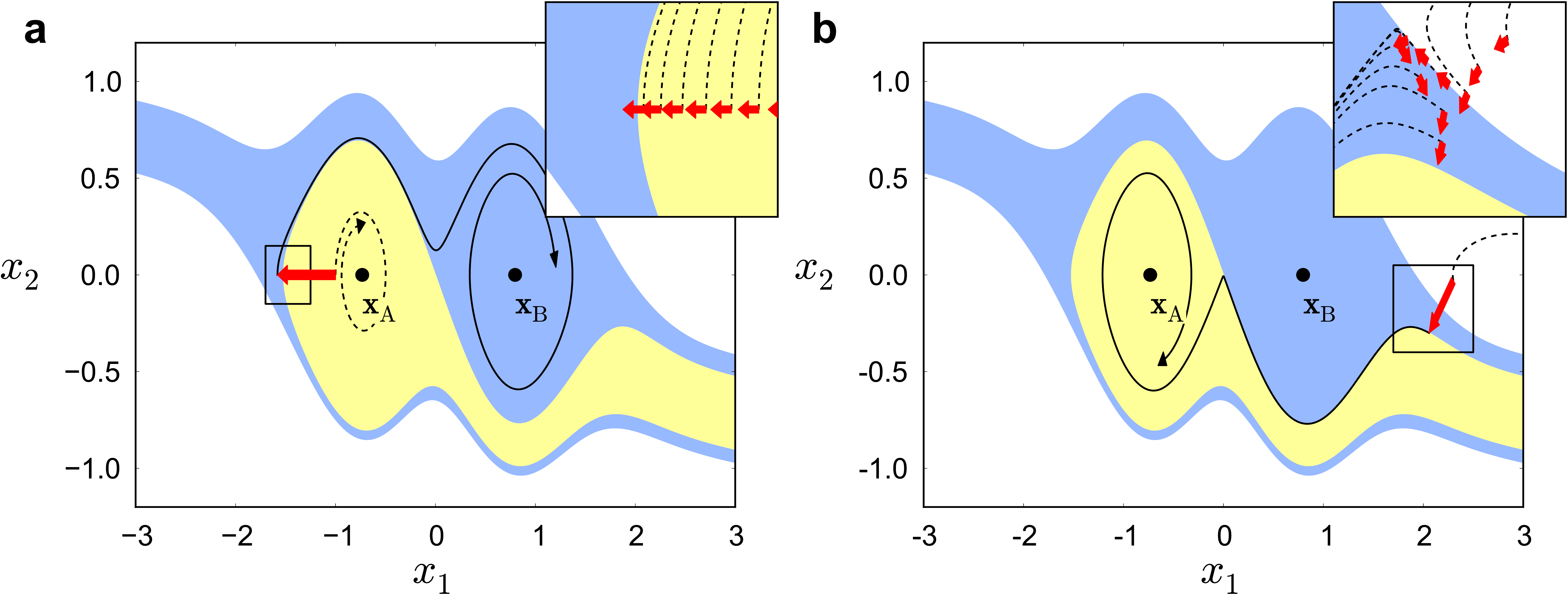}
\caption{{\bf Illustration of the control process in two dimensions.} Yellow and blue represent the basins of attraction of the stable states $\x_\mathrm{A}$ and $\x_\mathrm{B}$, respectively,  while white corresponds to unbounded orbits. ({\bf a}, {\bf b}) Iterative construction of the perturbation for an initial state in the basin of $\x_\mathrm{A}$  with $\x_\mathrm{B}$ as a target ({\bf a}), and for an initial state on the right side of both basins with $\x_\mathrm{A}$ as a target ({\bf b}).  Dashed and continuous lines indicate the original and controlled orbits, respectively. Red arrows indicate the full compensatory perturbations. Individual iterations of the process are shown in the insets (for clarity, not all iterations are included). \label{2d_example_fig}} 
\addcontentsline{toc}{subsection}{Supplementary Figure \thefigure}
\end{figure*}

\newpage
\clearpage
\begin{figure*}[th!]
\begin{center}
\includegraphics[angle=0,width=4.0in]{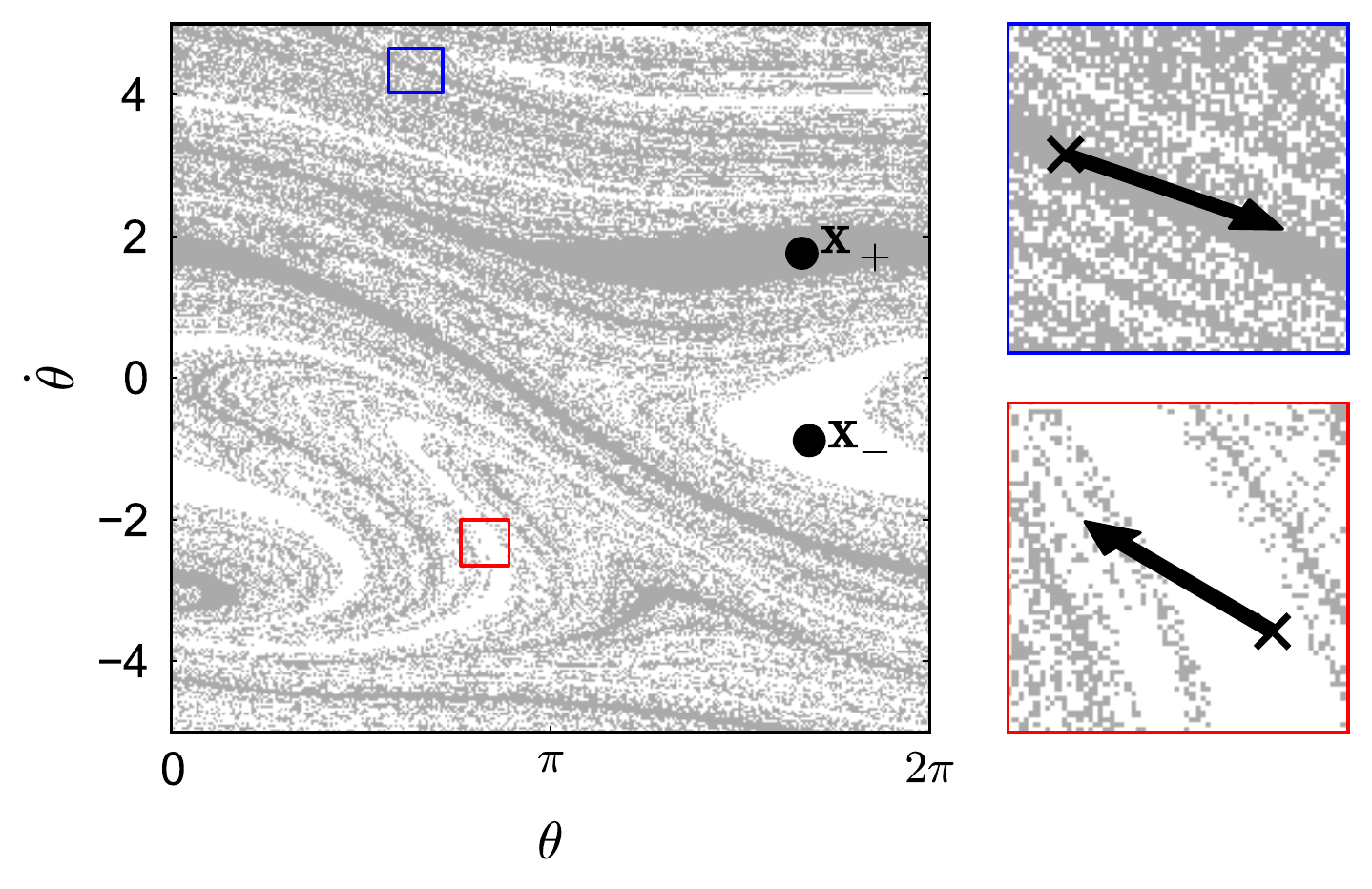}
\caption[\!]{{\bf Examples of compensatory perturbations in a system with fractal basins}. Section of the state space of eq.\ (\ref{fractal_system}) at $t = 0 \mod 2 \pi$. The basins of the clockwise ($\x_+$) and counterclockwise ($\x_-$) attractors are colored grey and white, respectively. The basins were calculated by sampling the pictured portion of the state
space at a resolution of $1,000$ points along each coordinate direction. The points at which the attractors strike the plane, which are
taken as the corresponding target states within our method, are marked with a red and blue $\times$, respectively. The right panels show examples of compensatory perturbations found by our method that take an initial condition (black $\times$) in the basin of $\x_-$ and move it to the basin of $\x_+$ (red), and vice versa (blue).
} \label{fractal_fig}
\end{center}
\addcontentsline{toc}{subsection}{Supplementary Figure \thefigure}
\end{figure*}

\newpage
\clearpage
\begin{figure*}[th!]
\begin{center}
\includegraphics[angle=0,width=4.0in]{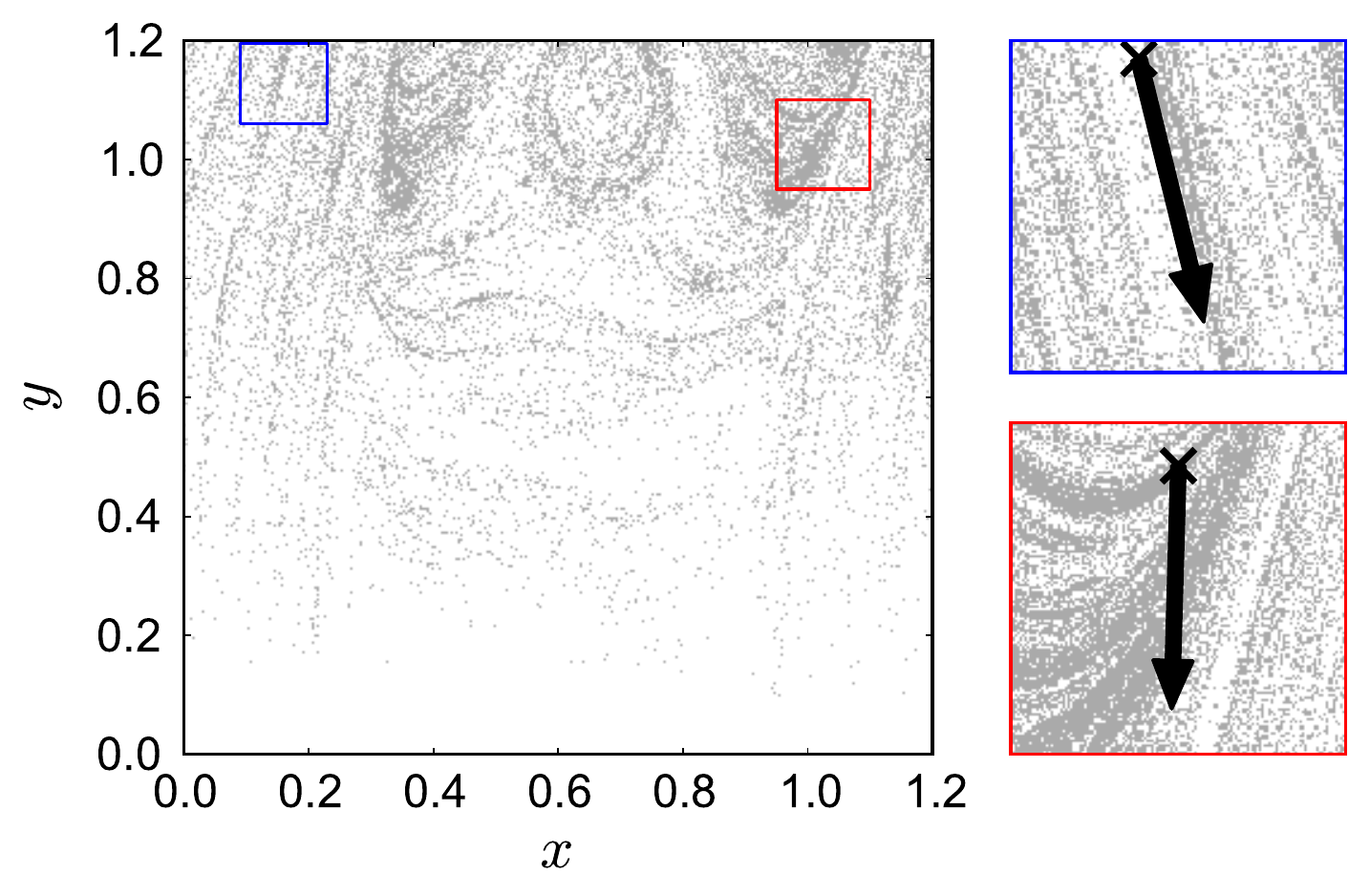}
\caption[\!]{{\bf Examples of compensatory perturbations in a system with riddled basins}. Section of the state space of eq.\ (\ref{riddled_system}) through $\dot{x} = \dot{y} = t \mod \frac{2\pi}{\omega} = 0$. Initial conditions corresponding to unbounded orbits ($\lvert y \rvert \rightarrow \infty$) are shaded in grey. The basin of attraction of the chaotic
attractor that lies in the subspace defined by $y = \dot{y} = 0$ appears in white. Initial conditions were sampled at a resolution of $1,000$ points along each coordinate direction. The right panels show two examples of compensatory perturbations found by our method that take initial conditions corresponding to unbounded orbits (black $\times$) and drive them into the basin of the chaotic attractor.
} \label{riddled_fig}
\end{center}
\addcontentsline{toc}{subsection}{Supplementary Figure \thefigure}
\end{figure*}

\newpage
\clearpage
\begin{figure*}[th!]
\centering
\includegraphics[width=16.0cm]{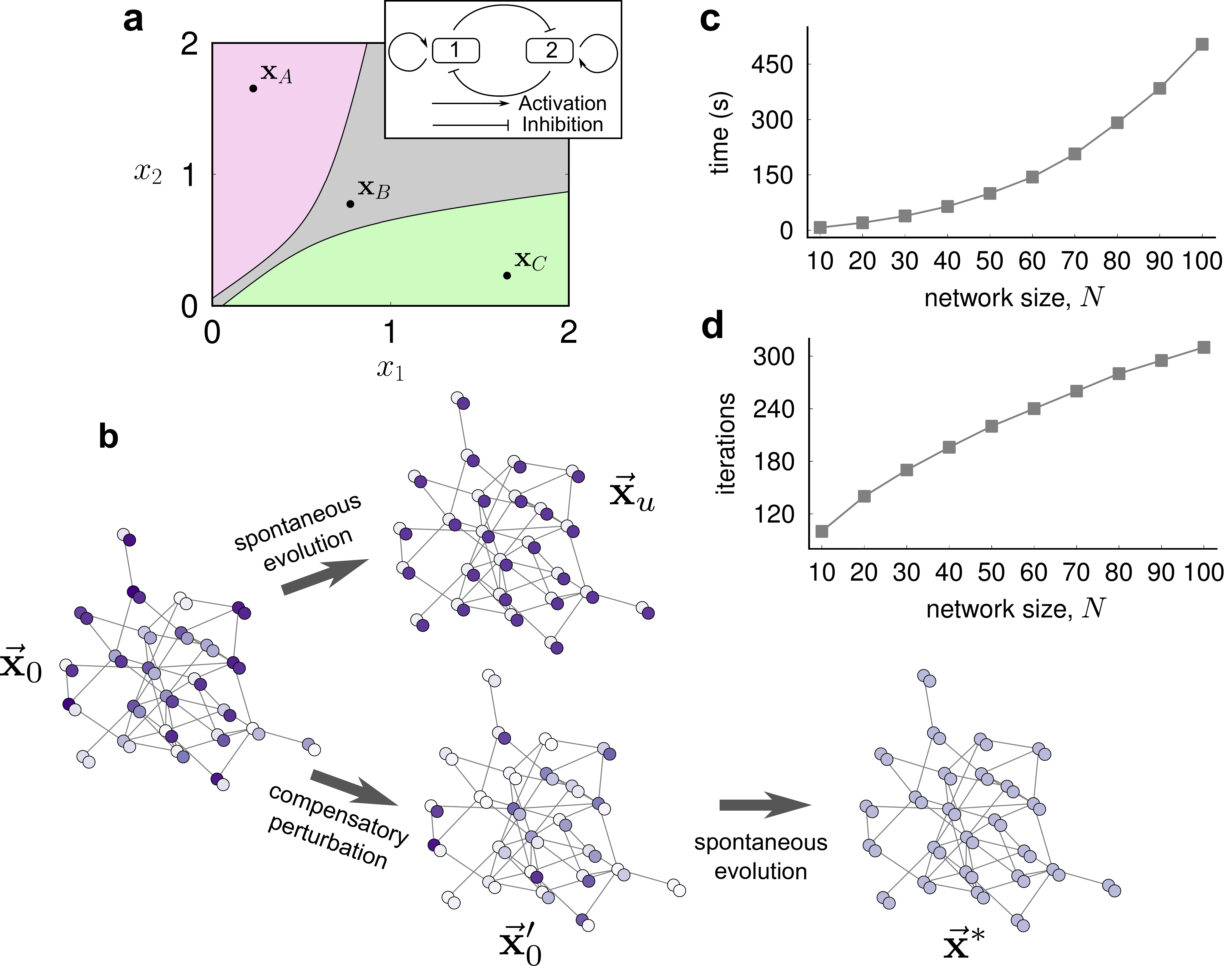}
\caption{{\bf Control of large random networks.}
({\bf a}) State space of the two-gene subnetwork represented in the inset, where the curves mark the boundaries between the basins of $\x_\mathrm{A}$, $\x_\mathrm{B}$, and $\x_\mathrm{C}$.  ({\bf b}) Illustration of compensatory perturbation on the genetic networks described by equation (\ref{syncdynamics}), where each node is a copy of the two-gene system. We are given an initial network state $\vec{\x}_0$ representing the expression levels of the $N$ gene pairs (color coded), and this state evolves to a stable state of the network $\vec{\x}_u$ (top path).  The goal is to knockdown one or more genes to reach a new state $\vec{\x}_0'$ that instead evolves to a  target stable state $\vec{\x}^*\neq \vec{\x}_u$ (bottom path). ({\bf c}, {\bf d}) Average computation time ({\bf c}) and average number of iterations ({\bf d}) required to control networks of $N$ nodes  with initial state $\vec{\x}_0 = \vec{\x}_\mathrm{A}$ and target $\vec{\x}^* = \vec{\x}_\mathrm{B}$, demonstrating the good scalability of the algorithm. Each point represents an average over $1,000$ independent random network realizations. 
\label{random_networks_fig}} 
\addcontentsline{toc}{subsection}{Supplementary Figure \thefigure}
\end{figure*}

\newpage
\clearpage
\begin{figure*}[th!]
\begin{center}
\includegraphics[angle=0,width=12.0cm]{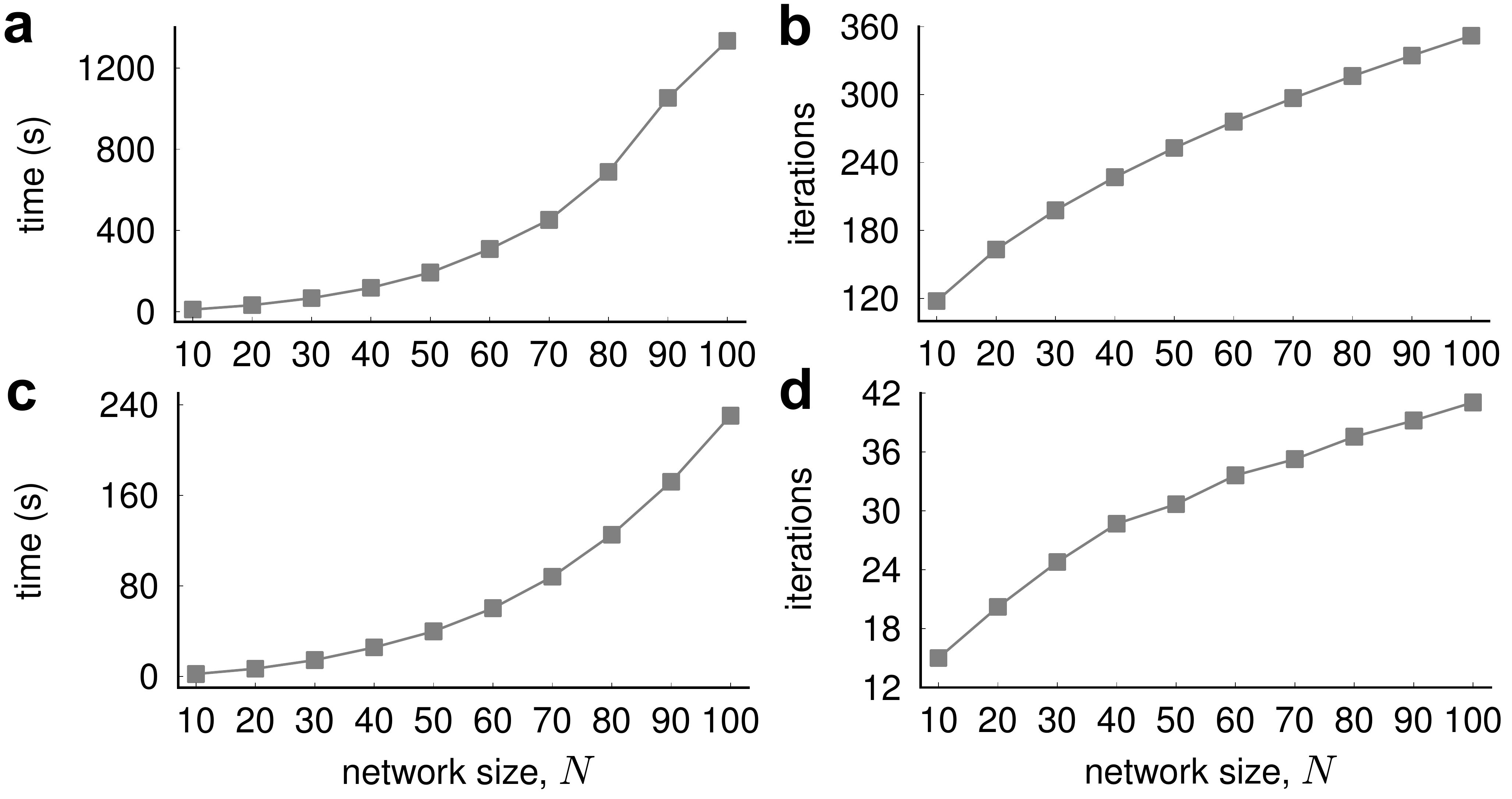}
\caption[]{{\bf Efficiency for alternate initial and target states.} Counterparts to Supplementary Figure \ref{random_networks_fig}c-d
for: ({\bf a}, {\bf b}) initial state $\vec{\x}_0 = \vec{\x}_\mathrm{A}$, target state $\vec{\x}^* = \vec{\x}_\mathrm{C}$; and 
({\bf c}, {\bf d}) initial state $\vec{\x}_0 = \vec{\x}_\mathrm{B}$, target state $\vec{\x}^* = \vec{\x}_\mathrm{A}$.
The other combinations of initial and target states involving $\vec{\x}_\mathrm{A}$, $\vec{\x}_\mathrm{B}$, and $\vec{\x}_\mathrm{C}$ 
follow from these ones and the one in Supplementary Figure \ref{random_networks_fig}c-d by symmetry.
The  approximately $N^{5/2}$
dependence of the computation time and sublinear dependence 
of the number of iterations are largely independent of the particular initial and
target states under consideration. 
\label{random_other_ics_fig}}
\addcontentsline{toc}{subsection}{Supplementary Figure \thefigure}
\end{center}
\end{figure*}

\newpage
\clearpage
\section*{Supplementary Table}
\addcontentsline{toc}{section}{Supplementary Table}
\setlength{\tabcolsep}{12pt}
\begin{table}[hb!]
\centering
\caption{{\bf Control procedure parameter values}. The relevant parameters are the integration time ($\tau$) and tolerance ($\kappa$) used to test convergence
to the target state, the limit on the number of iterations ($I$), the lower ($\epsilon_0$) and upper ($\epsilon_1$) bounds on the size of each incremental
perturbation, and the time window ($T$) over which the closest approach to the target is identified.}
\begin{small}
\begin{tabular}{lrrrrrr}
 & $\tau$ & $\kappa$ & $I$ & $\epsilon_0$ & $\epsilon_1$  & $T$ \\
\hline
T Cell signaling network     &  $10^3$ & $10^{-2}$ & $10^4$ & $10^{-3}$ 				& $10^{-2}$ 		& 5   \\
Associative memory network     & $2 \times 10^2$ & $10^{-2}$ & $10^4$ & $10^{-3}$ 				& $10^{-2}$ 		& 10  \\
New England power grid &  $10^3$ & $10^{-1}$ &  $10^4$ & $10^{-2}$ 				& $10^{-1}$ 		& 10  \\
2D potential system 	& $10^4$ & $10^{-2}$ & $10^3$ & $10^{-3}$					& $10^{-2}$ 		  & 10   \\
Random genetic networks	& $10^4$ & $10^{-2}$ & $10^4$ & $5 \times 10^{-3}$ 		& $5 \times 10^{-2}$ & 10   \\
\end{tabular}
\end{small}
\label{tableS1}
\addcontentsline{toc}{subsection}{Supplementary Table \thetable}
\end{table}

\newpage
\section*{Supplementary Methods}
\addcontentsline{toc}{section}{Supplementary Methods}

\subsection*{Termination criteria and control parameters.}
\addcontentsline{toc}{subsection}{Termination criteria and control parameters}

\noindent The control procedure is terminated if the updated initial condition attracts to within a distance $\kappa$ of the
target state within $\tau$ time units.  Otherwise,  we terminate the search if a compensatory perturbation is not found
after a fixed number $I$ of iterations. In general, this number should be of the order of $L$/$\epsilon_0$, 
where $L$ is the characteristic linear size of the feasible region. For each iteration, we use $T$ time units 
within the integration step that identifies $t_\mathrm{c}$, which was estimated based on the time to approach the undesirable stable state. 
As described in the main text, the parameters $\epsilon_0$ and $\epsilon_1$ define the minimum and maximum size
of the incremental perturbation within the optimization step, respectively.
Table  \ref{tableS1} lists the parameter values
used for each of the systems we have studied. 

In all systems we have studied, when a solution cannot be found, this is manifested in our method as an inability to move the orbit any closer to the target, which eventually leads to oscillations within the feasible region. The parameter $I$ is always taken to be large enough that this occurs before the 
iteration limit is reached. This also points to an alternative formulation of the termination criterion in which the procedure is terminated if it revisits an initial condition $\x_0' $ from a prior iteration within a distance less than $\epsilon_0$.

\subsection*{Construction of random genetic networks.}
\addcontentsline{toc}{subsection}{Construction of random genetic networks}
\noindent
The networks used in the Supplementary Discussion are grown starting with a $d$-node connected seed network, by iteratively attaching a new node and connecting this node to each pre-existing node $i$  with probability  $d\times P_i$, where $\sum_i P_i=1$. The connections are assumed to be unweighted and undirected. 
We reject any iteration resulting  in a degree-zero node, thereby ensuring that the final $N$-node network is connected and has average degree $>2d$ for large $N$. Networks are generated with uniform attachment probability, where $P_i= 1/N^{(q)}$ and $N^{(q)}$ is the number of nodes in the network at iteration $q$.
In our simulations, we focused on networks with $d=2$.

\subsection*{Drawbacks of alternative implementations.}
\addcontentsline{toc}{subsection}{Drawbacks of alternative implementations}
\noindent
The method introduced in the main text should be compared with a number of apparently simpler implementations that could, in principle, be used to search
the network state space for the purposes of finding a compensatory perturbation. For example, rather than keeping track of the  variational matrix ${\mathbf M}(\x_0,t)$, which requires the integration of $n^2$ additional  differential equations at every iteration, one could imagine using backward integration of a trajectory starting at $\x(t_\mathrm{c})+\delta \x(t_\mathrm{c})$ to identify a suitable initial  perturbation.  This alternative procedure suffers the critical drawback that, for a particular choice of $\delta \x(t_\mathrm{c})$ (magnitude and direction), it is not certain that the time-reversed orbit will ever strike the feasible region defined by  equation  (2) (main text). This is particularly so in realistic situations where only a fraction of the nodes are accessible to perturbations, resulting in a feasible region of measure zero in the full $n$-dimensional state space. 
	
Similar in spirit to our approach are standard shooting or ``shoot-and-look'' type methods, which seek to solve an initial value problem such that it satisfies a particular boundary condition (in this case, reaching the target at time $\tau$). These methods operate by making repeated adjustments to the initial condition $\x_0$, observing how those adjustments change the orbit at a later time, and then keeping those adjustments that move the orbit closer to the intended boundary value. The problem is that in each step, in order to ``look'' we must first ``shoot'', which requires an expensive integration of the system dynamics. Our approach circumvents this by calculating the mapping between {\it all} (small) initial perturbations and their images at later times (the variational matrix $M$), which allows us to determine the optimal ``shot'' at every iteration with only one integration. Similar problems afflict other techniques that involve random adjustment and then updating of data, such as simulated annealing. In the high-dimensional state spaces typical of real networks, each successful update of the initial condition within these methods simply requires too many trial sub-iterations, each of which entails evolving the system dynamics. 

\newpage
\section*{Supplementary Discussion}
\addcontentsline{toc}{section}{Supplementary Discussion}

\subsection*{Illustration of control procedure in two dimensions.}
\addcontentsline{toc}{subsection}{Illustration of control procedure in two dimensions}
It is instructive to apply the control procedure introduced in the main text to an example in two dimensions, where the basins of attraction (and hence the possible compensatory perturbations) can be explicitly calculated and visualized.  Supplementary Figure \ref{2d_example_fig} shows the state space of the system, which has two stable states, $\x_\mathrm{A}$ on the left and $\x_\mathrm{B}$ on the right. The system consists of a particle in one dimension under the influence of the potential $U(x_1)= \exp (-\gamma x_1^2)(bx_1^2+cx_1^3+dx_1^4)$ and frictional dissipation $\eta$,  where $\gamma=1$, $b=-1$, $c=-0.1$, $d=0.5$, $\eta=0.1$, and $x_2=\dot{x}_1$.  The method is illustrated for two different initial states under the constraint that admissible perturbations have to satisfy  $\x_0'\le \x_0$, i.e., one cannot increase  either variable. 

For the initial state in the basin of state $\x_\mathrm{A}$ (Supplementary Fig.~\ref{2d_example_fig}a), no admissible perturbation exists that can bring the system directly to the target $\x_\mathrm{B}$, on the right, since that would require increasing $x_1$. However, our iterative procedure builds an admissible perturbation vector that shifts the state of the system to a branch of the basin $\Omega(\x_\mathrm{B})$ lying on the left of that point. Then, from that instant on the autonomous evolution of the system will govern the trajectory's approach to the target $\x_\mathrm{B}$, on the right. This example illustrates how compensatory perturbations that move in a direction away from the  target---the only ones available under the given constraints---can be effective in controlling the system, and how they are identified by our method. The other example shown illustrates a case in which the perturbation to an initial state on the right crosses an intermediate basin, that of $\x_\mathrm{B}$, before it can reach the basin of the target, $\x_\mathrm{A}$ (Supplementary Fig.~\ref{2d_example_fig}b). The linear approximation fails at the crossing point, but convergence is nonetheless assured by the constraints imposed on $\delta\x_0$ (\emph{Methods}).
Our method is similarly capable of dealing with even more complicated basin structures, such as fractal or riddled basins (see {\it Effectiveness of control in the presence of fractal basins} and {\it Effectiveness of control in the presence of riddled basins} below). For a step-by-step animation of the control procedure in this system, please refer to Supplementary Movie.

\subsection*{Effectiveness of control in the presence of fractal basins.}
\addcontentsline{toc}{subsection}{Effectiveness of control in the presence of fractal basins}
\noindent
Systems with fractal basin boundaries may in principle be a challenge for our approach due to the possible existence of very long transients before the orbit comes near the relevant attractor. To test our approach's ability to cope with this scenario, we have applied it
to a well-known system exhibiting fractal basins, namely, a driven, dissipative oscillator, whose dynamics obey
\begin{equation}
\frac{d^2 \theta}{dt^2} + \nu \frac{d \theta}{dt} + \sin \theta = F \cos t, \label{fractal_system}
\end{equation}
where $\theta$ denotes the phase of the oscillator and the values of the damping parameter ($\nu = 0.1$) and forcing parameter ($F = 2.1$) are taken from [29]. We identify a unique state within the three-dimensional state space of this system by $\x = (x_1, x_2, x_3) = (\theta, \dot{\theta}, t)$, where the dot denotes a time derivative. This system posseses two periodic attractors each with period $2 \pi$ and distinguished by average clockwise or counterclockwise motion ($\langle \dot{\theta} \rangle$ positive or negative). We denote these attractors by $\x_+(t)$ and $\x_-(t)$, respectively. A section of the state space at $t=0 \mod 2 \pi$ is shown in Supplementary Figure \ref{fractal_fig}, with the two (fractal) basins of attraction shaded. Given an initial state $\x_0 = (\theta_0, \dot{\theta}_0, 0)$ in the basin of one attractor, we attempt to use our method to find a compensatory perturbation that places the system in the basin of the other attractor. We use as a target state the point on the other attractor at $t = 0 \mod 2 \pi$. Since we are primarily concerned with our method's ability (or inability) to find compensatory pertubations within the convoluted state space, the only constraint we apply is that eligible perturbations cannot adjust the artificial coordinate $x_3(0)$ that represents time. Note that since the dynamics are invariant under translation of either $\theta$ or $t$ by $2 \pi$, we measure differences along these coordinate directions modulo $2 \pi$ for the purposes of the distance metric used in our method. We use the conservative parameter choices of $\epsilon_0=10^{-4}$, $\epsilon_1=10^{-3}$, $\kappa = 10^{-3}$, $I=100,000$, $\tau = 100 \times 2\pi$, and $T = 2\pi$. The   integration time window $T$ is deliberately chosen to be short (only one period) as a test of the importance of long transients. In other words, our method will only consider a short initial portion of the orbit in deciding how to proceed at every iteration.

The right panels of Supplementary Figure \ref{fractal_fig} show the results of two control experiments, taking an initial point in the basin of $\x_-(t)$ and targeting $\x_+(t)$, and vice versa. Our method successfully finds states in the target basin in both cases, and with little apparent difficulty---only a few hundred iterations are required in either case, and we observe that the method makes more or less consistent progress moving the orbit closer to the target. While we have shown only two examples for clarity, these results are representative of $1,000$ initial states, chosen randomly from the phase plane of Supplementary Figure \ref{fractal_fig}, each of which we attempt to bring to the opposite basin. Remarkably, our method is successful in over $99\%$ of these cases.
We posit that long transients are not an issue because it is not critical that the uncontrolled orbit come ``near'' the target, in any absolute sense. Rather, all that is necessary is a point on the orbit at which progress can be made through an incremental perturbation. In the worst case, a long transient simply means one must evolve the system dynamics for a longer time to verify which attraction basin the current initial condition belongs to. 

\subsection*{Effectiveness of control in the presence of riddled basins.}
\addcontentsline{toc}{subsection}{Effectiveness of control in the presence of riddled basins}
\noindent
Another complicated basin structure that can arise in special cases is a so-called {\it riddled} basin, in which every point in the basin has points of a different basin arbitrarily closeby (the basin is ``riddled'' with holes). While this property could in principle pose problems for the identification of compensatory perturbations, we verified that our approach can perform quite well in such systems. This is the case because, as for other attractors, these basins too have non-zero measure in the state space. We used a most widely-known system with a riddled basin introduced in ref.~\citen{ott_1993}. The system consists a point particle of unit mass moving in two dimensions $\mathbf{r} = (x, y)$ under the influence of the potential $V(x, y) = (1-x^2)^2 + (x + \bar{x})y^2$, with dynamics governed by
\begin{equation}
\frac{d^2\mathbf{r}}{dt^2} = -\nabla V + \mathbf{e}_x f_0 \sin \omega t \mathbf - \nu \frac{d\mathbf{r}}{dt}, \label{riddled_system}
\end{equation}
where the  second and third terms represent the influences of a driving force and friction, respectively, and $\mathbf{e}_x$ is a unit vector in the $x$ direction. We represent a point within the five-dimensional state space by $\x = (x, \dot{x}, y, \dot{y}, t)$ and use the parameter values $f_0 = 2.3$, $\omega = 3.5$, $\bar{x} = 1.9$, and $\nu = 0.05$ given in ref.~\citen{ott_1993}. For these parameters, this system posseses a chaotic attractor $\x_\mathrm{A}(t)$ in the subspace defined by $y = \dot{y} = 0$, whose basin is riddled with initial conditions corresponding to unbounded orbits ($\lvert y \rvert \rightarrow \infty$). A ($x_0$, $y_0$) slice of the state space showing the riddled basin structure is depicted in Supplementary 
Figure \ref{riddled_fig} for $\dot{x} = \dot{y} = t = 0$.  Given an initial condition in this plane corresponding to an unbounded orbit (grey), we attempt to use our method to bring it to the basin of attraction of the chaotic attractor (white) by perturbing only the coordinates $x_0$ and $y_0$.  Since the desired attractor is an extended set in this case, we use a single point on the attractor (namely $(x, \dot{x}, t) = (-0.991, -1.254, 1000)$) as the target state within our method, which is not directly reachable by any eligible perturbation. Nonetheless, our method is $100 \%$ successful within a sample of 1,000 initial states selected randomly from the unbounded ``basin'' depicted in Supplementary Figure \ref{riddled_fig}. The parameters used for the control procedure in this analysis were $\epsilon_0=10^{-4}$, $\epsilon_1=10^{-3}$, $\kappa = 10^{-3}$, $I=100,000$, $\tau = 1000$, and $T = 10$.

\subsection*{Effectiveness and computational efficiency.}
\addcontentsline{toc}{subsection}{Effectiveness and computational efficiency}
\noindent In order to validate our procedure for identifying compensatory perturbations in networks, we consider networks of diffusively-coupled units---a case that has received much attention in the study of spontaneous synchronization\cite{pecora_prl_2008}.  We take as a base system the genetic regulatory subnetwork shown in Supplementary Figure
\ref{random_networks_fig}a (inset), consisting of  two genes wired in a circuit. The state of the system is determined by the expression levels of the genes, represented by the variables $x_1, x_2 \geq 0$. The associated dynamics obeys
\begin{align} 
\frac{dx_1}{dt} &= a_1 \frac{x_1^m}{x_1^m + S^m} + b_1 \frac{S^m}{x_2^m + S^m} - k_1 x_1 + f_1,  \label{twogene}  \\
\frac{dx_2}{dt} &= a_2 \frac{x_2^m}{x_2^m + S^m} + b_2 \frac{S^m}{x_1^m + S^m} - k_2 x_2 + f_2,  \label{twogene2}
\end{align}
where the first two terms for each gene capture the self-excitatory and mutually inhibitory interactions represented in Supplementary Figure \ref{random_networks_fig}a, respectively, while the final two terms represent linear decay ($k_{1,2}$) and basal activation ($f_{1,2}$) rates of the associated gene's expression. The parameter
$S$ represents the threshold above (below) which each gene is considered ``on'' (``off'').
While used here as a benchmark to test our computational approach, it is worth noting that models of this form have been employed to describe the transition between progenitor stem cells and differentiated cells\cite{Roeder_jtb_2006,Huang_DevBiol_2007,huang_biophysj_2009}, and that genetic ``switch'' circuits have been recognized as important motifs for the control of biochemical networks\cite{collins_2012}. For a wide range of parameters, this system exhibits three stable states: a state ($\x_\mathrm{B}$) characterized by comparable expression of both genes, and two states  ($\x_\mathrm{A}$ and $\x_\mathrm{C}$) characterized by the dominant expression of one of the genes.  The former corresponds to a stem cell state, and the latter correspond to two distinct differentiated cell types. Supplementary Figure  \ref{random_networks_fig} and subsequent results correspond to the symmetric choice of parameters $a_{1,2}=0.5$, $b_{1,2}=1$, $k_{1,2}=1$, $f_{1,2}=0.2$, $S = 0.5$, and $m=4$. We assume that compensatory perturbations are limited to decreases in gene expression, i.e., $\x_0'\le \x_0$. Our procedure applied to this system identifies compensatory perturbations between all three stable states (Supplementary Movie). 
 
We construct large networks by coupling multiple copies of the two-gene system described above (Supplementary Methods).
Such intercellular genetic networks may represent cells in  tissue or culture coupled by means of factors exchanged through their microenvironment or medium. 
Specifically, we assume that each copy of the genetic system in equations (\ref{twogene})-(\ref{twogene2}) can be treated as a node of the larger network. 
The dynamics of a network consisting of  $N$ such systems is then governed  by
\begin{equation} \label{syncdynamics}
\frac{d\x_i}{dt} = \vect{f}(\x_i) + \frac{\sigma}{d_i} \sum_{j=1}^N A_{ij} [\x_j  - \x_i],
\end{equation}
where $\dot{\x}_i = \vect{f}(\x_i)$ is the vectorial form of the dynamics of node $i$ as described by equations  (\ref{twogene})-(\ref{twogene2}), the parameter $\sigma>0$ is the overall coupling strength, and $d_i$ is the degree (number of connections) of node $i$. The structure of the network itself is encoded in the adjacency matrix  $A = (A_{ij})$.  We use $\vec{\x}=(\x_i)$ to denote the state of the network, with $\vec{\x}_\mathrm{A}$,  $\vec{\x}_\mathrm{B}$,  and $\vec{\x}_\mathrm{C}$ denoting the network states in which all nodes are at state $\x_\mathrm{A}$,  $\x_\mathrm{B}$,  and $\x_\mathrm{C}$, respectively.  The states $\vec{\x}_\mathrm{A}$, $\vec{\x}_\mathrm{B}$,  and $\vec{\x}_\mathrm{C}$ are fixed points of the full network dynamics in the $2N$-dimensional state space and, by arguments of structural stability, we can conclude they are also stable and have qualitatively similar basins of attraction along the coordinate planes $\x_i$ if the coupling strength $\sigma$ is weak.  While we focus on these three states,  it follows from the same arguments that in this regime there are $3^N\! - 3$ other  stable states in the network. Under such conditions, compensatory perturbations between $\vec{\x}_\mathrm{A}$,  $\vec{\x}_\mathrm{B}$,  and $\vec{\x}_\mathrm{C}$ are guaranteed to exist, and hence this class of networks can also serve as a benchmark to test the effectiveness and efficiency of our method in finding compensatory perturbations in systems with a large number of nodes.  

A general compensatory perturbation in this network is illustrated  in Supplementary Figure \ref{random_networks_fig}b, where different intensities indicate different node states. Applied to the initial state  $\vec{\x}_\mathrm{A}$ and target  $\vec{\x}_\mathrm{B}$ for $\sigma = 0.05$, the method is found to be effective in $100\%$ of the cases for the $10,000$ networks tested, with $N$ ranging from $10$ to $100$. Moreover,  the computation time and number of iterations for these tests confirm that our method is also computationally efficient for large networks.  Computation time grows polynomially with $N$ (Supplementary Fig.~\ref{random_networks_fig}c), as expected since each iteration requires the integration of  $O(N^2)$ equations and each equation can be integrated in $O(1)$ time as long as the average degree remains essentially constant, as is the case in many network models. The number of iterations grows as the square root of $N$ (Supplementary Fig.~\ref{random_networks_fig}d), in agreement with the $\sqrt{N}$ scaling of $\lvert \vec{\x}_\mathrm{A} - \vec{\x}_\mathrm{B} \rvert$ and of the distances between other invariant sets. This leads to the asymptotic scaling $N^{5/2}$ for the computation time. These properties are representative of other choices of initial and target states (Supplementary Fig.~\ref{random_other_ics_fig}).  
Note that our argument for the time complexity does not depend on the particular functional form of the dynamics nor on any parameter other than the dimension of the state space (which is usually proportional to the number $N$ of nodes, as assumed here).

\subsection*{Comparison with existing literature.}
\addcontentsline{toc}{subsection}{Comparison with existing literature}
\noindent
	Note that our approach is fundamentally different from those usually considered in control theory, both in terms of methods and applicability. To appreciate this, it is instructive to compare the problem addressed here---and the solution offered---to other important problems that fall under the broad umbrella of ``control''.  

	One well-developed meaning of ``control'' entails the optimization of specific system properties. Optimal control\cite{Athans2006}, for example, is based on identifying an admissible (time-dependent) control signal ${\mathbf u}(t)$ such that an orbit of the modified system $d{\mathbf x}/{dt}={\mathbf G}(\x, {\mathbf u}, t)$ will optimize a given cost functional $J(\x, \mathbf{u})$. In this representation, the discrete interventions we consider would take the form ${\mathbf u}(t) = {\mathbf u}_0 \delta (t - t_0)$, where $\mathbf{u}_0$ is then a compensatory perturbation to be determined. At first glance then, the discrete form of the controls we seek is the same as those used in impulse control\cite{kurzhanski_2008}. But we stress that here, the challenge is not to identify an ``optimal'' solution but a {\it valid} solution in the first place ({\it i.e.}, an eligible point $\x_0 + \mathbf{u}_0$ inside the target's basin of attraction). This is a goal that cannot be easily guided by global optimization of any particular ``cost'' in a computationally tractable way, nor formulated simply as one or more closed form (in)equality constraints. For this reason, the extensive and well-developed machinery of optimal control, impulse control, model predictive control, and related methodologies unfortunately cannot be directly applied to solve the problem considered here.
	
Another important sense of  ``control'' concerns the stabilization of otherwise unstable (and therefore uninhabitable) states. Control of chaos\cite{ott95}, for instance, can be used to convert a chaotic trajectory into a periodic one, and is based on the continuous application of unconstrained small time-dependent perturbations to align the stable manifold of an unstable periodic orbit with the trajectory of the system. Similar methods have been applied, for example, to stabilize desirable periodic behaviors in models of cardiac activity\cite{brandt_1997}. Conversely, techniques have also been developed to destroy undesirable attractors by using an appropriate modulation of the system parameters\cite{pisarchik_prl_2000, pisarchik_pre_2001}. Here, however, we do not seek to create or annihilate a stable state, but rather to bring the system to an existing state that is already stable. Moreover, we seek to do this using one (or few) constrained finite-size perturbations to the system {\it state} (rather than the dynamical system itself), and these perturbations are forecast-based rather than feedback-based.
	
More similar to the sense of ``control'' that we consider here is the method of targeting\cite{bollt2003}, which can be used to facilitate the approach to a desired orbit. But like control of chaos, this method generally applies when one wishes to  move within the same ergodic component rather than to move between different basins of attraction as we do here. And although a few methods have been developed to accomplish such transitions in multiple-attractor systems, they generally require prior knowledge about features of the network state space either directly\cite{jackson_1990} or indirectly via auxiliary techniques such as Lyapunov functions\cite{richter_2002}. This unfortunately limits their applicability to low-dimensional systems and special cases for which such information is available. Indeed, the central challenge that our approach seeks to overcome is the identification of control interventions without reliance on details of global properties of the state space of the network in question.

A closer precedent to our work is ref.~\citen{sagar_natcom_2011},  where examples of compensatory perturbations were provided for food-web networks. They were identified, however, by seeking to make the current state of the system similar to a desired state.  Such heuristics do not directly account for the subsequent time evolution of the perturbed orbit (i.e., the control perturbation is not forecast based) and they do not take systematic advantage of the critical role of basins of attraction, which, as demonstrated here, allow control even when the target itself is not directly accessible. It should also be noted that while our approach makes use of constrained optimization\cite{Bazaraa2006}, the question at hand cannot be formulated as a simple optimization problem in terms of an aggregated objective function, such as the number of active nodes. Maximizing this number by ordinary means can lead to local minima or transient solutions that then fall back to asymptotic states with a larger number of inactive nodes.  The identification of stable states that enjoy the desired properties is thus an important step in our formulation of the problem.

None of this is to suggest that the other methodologies discussed in this section are in any way deficient. As we have noted, they simply address different problems from the one considered here. Our formulation of the control problem can nonetheless benefit from existing techniques. Once a compensatory perturbation has been found, existing control and optimization methods can be used to modify this solution so as to optimize specific properties. For example, it might be of interest to identify a solution that, among all eligible ones, brings the system to the target the fastest. Incidentally, for lying further inside the target basin of attraction, such solutions have the remarkable property of being resilient against noise and parameter uncertainty (see {\it Effects of stochasticity} and {\it Effects of parameter uncertainty} below, Supplementary Figs.~\ref{noise_schematic_fig}-\ref{network_params_fig}). These possibilities underscore the robustness and versatility of our core approach. We emphasize, however, that alternative compensatory perturbations can only be found with existing methods once at least one eligible state inside the target basin has already been identified. 

\subsection*{Effects of stochasticity.}
\addcontentsline{toc}{subsection}{Effects of stochasticity}
\noindent 
In our formulation of the network control problem and our procedure for
identifying compensatory perturbations, we have assumed the network dynamics is deterministic. In real systems,
however, there will often be noise in the dynamics. This raises the question of whether our approach
can be used to effectively control systems with a stochastic component.
	To address this question, we revisit here the compensatory perturbations predicted with our method on deterministic models,
and test their effectiveness in a noisy version of the same system. We do this for {\it i}) the mechanical example system presented in Supplementary 
Figure \ref{2d_example_fig}, and {\it ii}) 
the genetic network model with the initial/target states considered for Supplementary Figure \ref{random_networks_fig}. To model the effects of 
stochasticity, we include an additive noise term in the dynamics of each state variable, yielding the stochastic differential equation 
$d\x/dt = \mathbf{F}(\x) + \mathbf{\xi}(t)$ for the new dynamics. Following the notation in the main text, $\mathbf{F}$ is the $n$-dimensional deterministic dynamics, 
and $\mathbf{\xi}(t)$ is a vector of $n$ independent Gaussian white noise processes, each with mean $0$ and r.m.s.\ amplitude $s$ 
(which quantifies the strength of the noise). Starting from a state $\x_0'$ reached by a putative compensatory perturbation, we generate
many independent realizations of the noisy dynamics using a stochastic Runge-Kutta scheme \cite{kloeden_1998} to determine the
probability that the noisy system reaches the target.

Before we proceed with a numerical experiment, we must appropriately define the notions of ``stability'' and  ``attractors''
in the presence of stochasticity, since with the addition of noise fixed points are no longer strictly fixed. We consider
a noisy orbit to reach the target if, after the usual
integration time limit $\tau$ used to test convergence to the target in the deterministic case (Supplementary Methods),
the mean position $\langle \x(t) \rangle$ over an additional $1,000$ time units falls within a ball of radius $r$ around the target.
We then consider noise strengths $s$ up to a maximum $s_{\max}$, where the threshold $s_{\max}$
is defined such that at this noise level the standard deviation of the noisy orbit around the target in any direction
is at most $r$. Rigorously, we consider the stochastic dynamics near the target state $\x^*$, given by $d\x/dt = A \cdot \x + s\,\mathbf{\xi}(t)$,
where $A = D_{\x} \mathbf{F} \vert_{\x^*}$ is the Jacobian matrix of $\mathbf{F}$ evaluated at the target state.
Over long times, the maximum expected variance of this process along any coordinate direction is $\sim s^2/(2 \lvert \lambda_{\min} \rvert)$, where $s$
is the noise strength as defined above and $\lambda_{\min}$ is the real part of the eigenvalue of $A$ with smallest real part magnitude. Thus, our criterion
for the noise strength $s_{\max}$ that defines a fuzzy ``ball of stability'' of radius $r$ around the target is $s_{\max} = \sqrt{2 r  \lvert \lambda_{\min} \rvert}$.
We use $r = 0.1$ for the systems in this section, which yields $s_{\max} \approx 0.02$ and $s_{\max} \approx 0.03$ for the two-dimensional
mechanical system and random genetic network systems, respectively. 
	
	Supplementary Figure \ref{noise_schematic_fig}a-b illustrates the effect of noise for the compensatory perturbations presented in Supplementary Figure 
\ref{2d_example_fig}a-b, respectively.  Although the concept of a ``basin of attraction'' is not absolute in the presence of noise, it is instructive to interpret the effects of noise
in terms of how it might ``kick'' the orbit between the deterministically-defined basins---regions where the mean dynamics tends toward
one attractor or the other. Indeed, because our algorithm declares success after finding a point $\x_0'$ just inside the target's 
basin of attraction, it is possible for the noisy trajectory to wander back across the boundary into the basin of the other attractor (red curves).
But there is nothing in our formulation of the control problem that dictates we must perturb the system to any particular point
$\x_0'$; in general, there will exist other eligible states {\it further inside} the target's basin of attraction. By 
instead choosing one of those points, $\x_0''$, as the endpoint of our compensatory perturbation, the resulting orbit is
far more likely to stay within the deterministic target basin and as a result, far more likely to reach 
the target (grey curves). 

	Systematically, we choose the modified perturbed state $\x_0''$ by starting from $\x_0'' = \x_0'$ and minimizing the time $T(\x_0'')$ it takes the
deterministic orbit resulting from $\x_0''$  to reach a neighborhood of the target, namely, the same criterion used for attraction in the main text.
Any such state $\x_0''$ must also comply with the given constraints on the eligible compensatory perturbations defined by $\mathbf{g}(\x_0, \x_0'')$ and $\mathbf{h}(\x_0, \x_0'')$. 
This is done according to the nonlinear programming problem
\begin{eqnarray} \label{nlp-SI}
\textrm{min}_{\x_0''}  & & T(\x_0'') \nonumber \\
\textrm{s.t.} & & \mathbf{g}(\x_0, \x_0'') \leq 0 \label{ieq_constraint-SI} \\
& & \mathbf{h}(\x_0, \x_0'') =  0. \nonumber \label{eq_constraint-SI}
\end{eqnarray}
This problem can be solved by existing methods, and we implement it using Sequential Quadratic Programming (Methods) immediately 
after the application of our control procedure that identifies $\x_0'$. Since this problem is a single optimization of 
a continuous and well-defined objective function, this additional step is not costly. But note that solving the problem given by
equation (\ref{nlp-SI}) in such a straightforward way is possible only because we have solution $\x_0'$ that belongs to the basin of the target
and hence reaches the vicinity of the target in finite time. 

	Supplementary Figure \ref{noise_success_fig}a-b is a numerical demonstration of the effectiveness of the above procedure, for the respective control scenarios represented 
by the initial/target state combinations in Supplementary Figure \ref{2d_example_fig}a-b.  As the strength of the noise term is increased, the success rate of the original compensatory perturbations in both scenarios quickly drops to $\approx 50$-$60\%$ (red). As shown in Supplementary Figure \ref{noise_schematic_fig}, this can be attributed to the initial state's proximity to the boundary between the basins of $\x_\mathrm{A}$ and $\x_\mathrm{B}$, which allows noise to knock the orbit back and forth between the attraction basins. (Because the state $\x_0'$ {\it is} slightly inside the target basin, this process is biased, yielding a success rate above $50\%$). Nevertheless,
the slight modification described above yields initial conditions $\x_0''$ that are significantly more likely to reach their respective targets at 
all noise levels (grey), with success rates of $\approx 70\%$ (Supplementary Fig.~\ref{noise_success_fig}a) and $\approx 90\%$ 
(Supplementary Fig.~\ref{noise_success_fig}b) at the maximum noise strength. This is 
even more pronounced in the network case (Supplementary Fig.~\ref{noise_success_fig}c-d), where the effectiveness of the original compensatory perturbations degrades quickly 
with increasing noise, particularly in larger networks. This precipitous drop in the success rate is a consequence of the large number of attractors ($3^N$) in this model, 
which makes the effect of noise near the corresponding basin boundaries analogous to flipping a
many-sided coin to determine the ultimate fate of the network. These seemingly prohibitive odds highlight the resilience offered by our formulation
of the network control problem. By moving the system further inside the desired basin, namely to $\x_0''$, the modified interventions
ensure that the system will reach the target in nearly $100\%$ of cases up to near the maximum noise strength, even when the network size is
increased.

\subsection*{Effects of parameter uncertainty.}
\addcontentsline{toc}{subsection}{Effects of parameter uncertainty}
\noindent 
To formalize our approach in terms of the system's state space, we have assumed a model describing the dynamics
of the system under study. Such a model usually contains a (potentially large) number of parameters, and in practice the values of these
parameters may not be known precisely. A potential problem confronting our approach is thus that, when
predicted using an imperfect model, a compensatory perturbation may place the real system
in a different basin of attraction than the one intended.

To address this possibility, we performed the following analysis. We consider the genetic network model as used in Supplementary Figure \ref{random_networks_fig},  with weak coupling and the respective initial and target states  taken to be $\vec{\x}_0 = \vec{\x}_\mathrm{A}$ and $\vec{\x}^* = \vec{\x}_\mathrm{B}$. For all nodes, we use the nominal values of the parameters $a_{1,2}$, $b_{1,2}$, $k_{1,2}$, $f_{1,2}$, $S$, and $m$ presented in the main text. Given a network with these parameters, we predict two interventions: the original compensatory perturbation $\x_0 \rightarrow \x_0'$ found by our method, and a modified perturbation $\x_0 \rightarrow \x_0''$ given as a solution to the problem described by equation (\ref{nlp-SI}); like in our study of the effects of stochasticity above, $\x_0''$ is by design further inside the predicted target basin. 
Now, suppose that the nominal parameter values do not accurately represent the real system being modeled, and that the 
actual parameters lie somewhere in a window of uncertainty about their nominal values, which we take to be $\pm 5\%$ in our numerical experiments. Furthermore, suppose each of the six aforementioned parameters is allowed to vary in this way {\it independently} for each node, meaning that the coupled subsystems are no longer identical. We then test whether the predicted perturbations still drive the system to the target when the parameters have been altered to the actual values. We calculate the success rate of each perturbation across a number of these alternate parameter sets, chosen randomly within the range defined above. Supplementary Figure \ref{network_params_fig}a-b shows the results for networks of sizes $10$ and $20$, respectively. Although the original perturbation often fails to bring the
system to the target when the actual parameters differ from their nominal values (red bars), the modified perturbations succeed nearly $100\%$ of the time (grey bars), despite having been predicted based on a model that is ostensibly ``wrong''. 

Note that a change in the system parameters usually induces a change in the system's state space as well, including
the location, stability, and very existence of the fixed points. Thus, in all networks considered above, the target
state is displaced from its location in the imperfect model. We find the appropriate target state based on Newton's method applied
to the dynamics of the exact model,
starting from the target state determined in the imperfect model as an initial guess.  Any parameter assignment for which this procedure does not yield a stable fixed
point in the positive orthant for the new target is rejected. This is a minor restriction, especially in light of the fact that
most models are construed to faithfully reproduce the equilibria of interest. Supplementary Figure \ref{network_params_fig}c-d
shows the distribution of the amount the target moves  for the eligible parameter choices within the uncertainty
window we consider.

Here, as for the effect of noise considered above, we have shown that there is a simple, computationally inexpensive way 
to make the approach resilient against imperfections in the modeling. Both problems are addressed by a slight adjustment to our basic method, namely, moving the system further inside the predicted target basin by systematically minimizing the time taken to reach the target state. We emphasize that this is possible
because our formulation of the network control problem is based on taking advantage of extended features of the state space, such as basins of 
attraction. In this way, the challenges posed by stochasticity and parameter uncertainty highlight robustness as a fundamental strength of our approach.

\let\oldthebibliography=\thebibliography
\let\oldendthebibliography=\endthebibliography
\renewenvironment{thebibliography}[1]{%
    \oldthebibliography{#1}%
    \setcounter{enumiv}{46}%
}{\oldendthebibliography}
\renewcommand{\refname}{Supplementary References}

\end{document}